\documentclass[12pt]{article}
\usepackage{bbm}
\usepackage{epsfig}
\usepackage{array}
\usepackage{float} 
\usepackage{amsmath} 
\usepackage{amssymb}
\usepackage{amsfonts}
\usepackage{amscd}
\usepackage{bbold}

\usepackage{a4}

\usepackage{a4wide}
\usepackage{wasysym}

\def\be{\begin{equation}}
\def\ee{\end{equation}}
\def\gs{\mathrel{
   \rlap{\raise 0.511ex \hbox{$>$}}{\lower 0.511ex \hbox{$\sim$}}}}
\def\ls{\mathrel{
   \rlap{\raise 0.511ex \hbox{$<$}}{\lower 0.511ex \hbox{$\sim$}}}}

\newcommand{\ba}{\begin{array}{c}}
\newcommand{\baz}{\begin{array}{cc}}
\newcommand{\bad}{\begin{array}{ccc}}
\newcommand{\bea}{\begin{equation} \begin{array}{c}}
\newcommand{\eea}{ \end{array} \end{equation}}
\newcommand{\ea}{\end{array}}
\newcommand{\D}{\displaystyle}
\newcommand{\dms}{\mbox{$\Delta m^2_{\odot}$}}
\newcommand{\dma}{\mbox{$\Delta m^2_{\rm A}$}}
\newcommand{\meff}{\mbox{$\langle m \rangle$}}
\newcommand{\eV}{\mbox{ eV}}

\newcommand{\Eqref}[1]{Eq.\ \eqref{#1}}

\newcommand{\Cl}[1]{\mathcal{C} _{#1}}

\newcommand{\Ord}[2]{\; ^{\circ} \mathrm{#1}_{#2}  \;}
\newcommand{\OrdCl}[1]{\; ^{\circ} \mathcal{C} _{#1} \;}
\newcommand{\Groupname}[2]{$ {#1} _{#2} $}
\newcommand{\Rep}[1]{\underline{\mbox{\textbf{#1}}}}
\newcommand{\MoreRep}[2]{\underline{\mbox{\textbf{#1}}} _{\mbox{\textbf{#2}}}}
\newcommand{\Yuk}[1]{\kappa _{#1} \,}
\newcommand{\Order}[1]{\mathcal{O}(#1)}

\begin{document}

\title{
\hfill {\small TUM-HEP-582/05}\\
\vspace{-0.3cm}
\hfill {\small hep-ph/0503143} 
\vskip 0.7cm
\bf 
Minimal Mass Matrices for Dirac Neutrinos
}
\author{
Claudia Hagedorn\thanks{email: \tt claudia$\_\,$hagedorn@ph.tum.de}~~~and~~
Werner Rodejohann\thanks{email: \tt werner$\_\,$rodejohann@ph.tum.de}
\\ \\
{\normalsize \it Physik--Department, Technische Universit\"at M\"unchen,}\\
{\normalsize \it  James--Franck--Strasse, D--85748 Garching, Germany}
}
\date{}
\maketitle
\thispagestyle{empty}
\vspace{-0.8cm}
\begin{abstract}
\noindent 
We consider the possibility of neutrinos being Dirac particles and study 
minimal mass matrices with as much zero entries as 
possible. We find that up to 5 zero entries are allowed. 
Those matrices predict one vanishing mass state, $CP$ conservation and 
$U_{e3}$ either zero or proportional to $R$, where $R$ 
is the ratio of the solar and atmospheric $\Delta m^2$. 
Matrices containing 4 zeros can be classified in categories 
predicting $U_{e3} = 0$, $U_{e3} \neq 0$ but no $CP$ violation or  
$|U_{e3}| \neq 0$ and possible $CP$ violation. 
Some cases allow to set constraints on the neutrino masses. 
The characteristic value of $U_{e3}$ capable of distinguishing some of 
the cases with non--trivial phenomenological consequences is about 
$R/2~ \sin 2 \theta_{12}$.  
Matrices containing 3 and less zero entries imply (with a few exceptions) 
no correlation for the observables. 
We outline models leading to the textures based on the 
Froggatt--Nielsen mechanism or the non--Abelian discrete symmetry 
$D_4 \times Z_2$.

\end{abstract}

\newpage

\section{\label{sec:intro}Introduction}
Recent years witnessed a dramatic increase of our knowledge about the 
properties of neutrinos (for recent reviews, see for instance 
\cite{reviews}). The experimental evidence showed 
that at least two neutrinos possess 
a non--vanishing rest mass, whose 
value is bounded from above by roughly 1 eV \cite{APScosmo,APS0vbb}. 
This implied smallness of neutrino masses is commonly attributed to the 
see--saw mechanism \cite{seesaw}, which explains the tiny neutrino 
masses through their inverse proportionality to the mass scale of heavy 
Majorana neutrinos. 
In addition, the light neutrinos are predicted to be of Majorana type.  
The smoking--gun signature of this property 
would be the observation of neutrinoless double beta decay \cite{APS0vbb}. 
There is currently no incontrovertible evidence for this 
process (see the discussion in \cite{APS0vbb}) and therefore 
the possibility that neutrinos are Dirac particles, 
just like all the other known fermions, has to be 
considered as well. 
If neutrinos are indeed Dirac particles, their mass term is given just by 
the coupling of the lepton doublets with the Higgs doublet and 
a right--handed neutrino Standard Model singlet state. 
One should then explain however 
why the corresponding Yukawa coupling is so much smaller than the 
Yukawa coupling for the other fermions. 
There are several possibilities for this to arise. 
All have in common that effectively 
there is an additional $U(1)_{\rm B - L}$ symmetry which 
forbids a Majorana mass term. For instance: 
\begin{itemize}
\item in theories with large extra dimensions \cite{large} the 
effective Yukawa coupling is determined by the overlap of the wave functions 
of the SM particles (residing in the usual 3 + 1 dimensions) with the 
wave functions of the right--handed singlet neutrinos. 
The latter propagate in additional  
dimensions (the bulk) and a volume suppression of 
order $M_F/M_{\rm Pl}$ occurs, where $M_{\rm Pl}$ is the Planck scale 
and $M_F$ the fundamental gravity scale, possibly as low as 
10 TeV\@. Similar volume suppression can occur in models with 
warped extra dimensions or ``fat brane'' models \cite{otherED}.  
For more details of these possibilities, see the recent reviews in 
\cite{ays};   
\item in supersymmetric frameworks the suppression factor might be 
given by the SUSY breaking scale divided by the Planck scale in 
scenarios with anomaly mediated supersymmetry breaking \cite{mura}. 
For other supersymmetry or supergravity models of Dirac neutrino 
masses, see, e.g., \cite{sugraothers}; 
\item in superstring models it is possible to obtain small 
neutrino masses through a hierarchy of vacuum expectation values 
of SM singlets \cite{mova}. In this respect it is interesting to 
note that in string theories it seems difficult to obtain a simple 
see--saw mechanism \cite{string};  
\item if neutrinos are massless at tree level, small Dirac masses 
might be generated through radiative mechanisms, so that the 
smallness is explained by loop suppression \cite{loops};  
\item other approaches to generate Dirac neutrino masses are discussed in 
Refs.\ \cite{other}. 
\end{itemize}
A common argument in favor of Majorana neutrinos is that the 
implied lepton number violation can be used to generate the 
baryon asymmetry of the universe via the leptogenesis mechanism 
\cite{lepto}. However, even with Dirac neutrinos it is possible to 
achieve this \cite{DiracYB,DiracYB1}. 
We therefore conclude that there is some motivation 
for studying Dirac neutrinos.\\ 

\noindent Once the Dirac nature of neutrinos is accepted, one can ask 
immediately what the structure of the neutrino mass matrix $m_\nu$ 
looks like. In particular, minimal mass matrices are of 
interest since they usually 
imply interesting and testable correlations between the observables.  
For Majorana neutrinos, there has been some interest in texture zeros of 
$m_\nu$ \cite{2zeros0}. It has been found that only two independent 
zero entries are allowed by current data \cite{2zeros0}. 
Models leading to the successful textures \cite{2zeroseesaw} as well 
as renormalization effects \cite{2zerosRGE} have been studied. 
Here we perform a similar analysis for Dirac neutrino mass matrices. 
Since Dirac mass matrices are --- in contrast to 
Majorana mass matrices --- in general not symmetric, the results are 
expected to differ. We shall find that one is able to put up to 
5 zero entries in a Dirac neutrino mass matrix. Those cases predict one 
vanishing neutrino mass, $CP$ conservation and small or zero 
$U_{e3}$. Already for matrices with 4 zeros there are successful cases 
in which there is no correlation between the observables. 
Some of them display however a correlation different from the 5 zero 
cases. 

\noindent We also outline models based on the discrete 
symmetry group $D_4 \times Z_2$ or on the Froggatt--Nielsen mechanism, 
which allow 
in principle to reproduce the textures under consideration.\\

\noindent The paper is structured as follows: 
Section \ref{sec:gen} contains several general considerations about 
Dirac mass matrices, their form as implied by current data and comment on 
the possibility to determine the nature of the neutrino. 
Then we perform a systematic search  for zero 
textures for Dirac neutrinos in Section \ref{sec:min}, discussing the cases 
of 5, 4 and 3 or less zeros. We comment on the implications of the Dirac 
mass matrices being symmetric. Models based on the 
flavor symmetry $D_4 \times Z_2$ and on the 
Froggatt--Nielsen mechanism are outlined 
in Section \ref{sec:models}, before we conclude in Section \ref{sec:concl}.

\section{\label{sec:gen}Dirac Neutrino Mass Matrices: General Considerations}

For Dirac neutrinos, the PMNS \cite{PMNS} matrix can be parametrized as
\bea \label{eq:Upara}
U = \left( \bad 
c_{12} c_{13} & s_{12} c_{13} & s_{13} \, e^{i \delta} 
\\[0.2cm] 
-s_{12} c_{23} - c_{12} s_{23} s_{13} \, e^{-i \delta}
& c_{12} c_{23} - s_{12} s_{23} s_{13}\, e^{-i \delta}
& s_{23} c_{13} \\[0.2cm] 
s_{12} s_{23} - c_{12} c_{23} s_{13} \, e^{-i \delta}
& 
- c_{12} s_{23} - s_{12} c_{23} s_{13}\, e^{-i \delta}
& c_{23} c_{13}\\ 
               \ea   \right) 
\, , 
\eea
where $c_{ij}= \cos \theta_{ij}$, $s_{ij} = \sin \theta_{ij}$ and
$\delta$ denotes the $CP$ phase, sometimes called ``Dirac--phase''. 
Global analyzes of current data allow for the following values of the 
observables \cite{valle}: 
\bea \label{eq:valle}
\sin^2 \theta_{12} = 0.30^{+0.04}_{-0.05}~~~,~~~
\sin^2 \theta_{23} = 0.50^{+0.14}_{-0.12}~~~,~~~
\sin^2 \theta_{13} = |U_{e3}|^2 \le 0.031 \\[0.3cm]
\Delta m^2_{21} \equiv \dms = 
\left(7.9^{+0.6}_{-0.6} \right) \cdot 10^{-5} ~ {\rm eV}^2~~~,~~~
\left|\Delta m^2_{31}\right| \equiv \dma = 
\left(2.2 ^{+0.7}_{-0.5}\right) \cdot 10^{-3} ~ {\rm eV}^2\\[0.3cm]
\Rightarrow R \equiv \frac{\dms}{\dma} = 0.036^{+0.014}_{-0.011} ~, 
\eea 
where $\Delta m_{ij} ^{2}$ denotes $m_i^2 - m_j^2$. 
We gave the best--fit values and the allowed 
$2\sigma$ ranges. Regarding the mass ordering, both the normal 
and the inverted ordering can be accommodated, corresponding to the cases 
$m_3 > m_2 > m_1$ and $m_2 > m_1 > m_3$, respectively. 
The extreme cases are the normal hierarchy (NH), which is defined 
as $m_3^2 \simeq \dma \gg \dms \simeq m_2^2 \gg m_1^2$, and the inverted 
hierarchy (IH), which is 
defined as $m_2^2 \simeq m_1^2 \simeq \dma \gg m_3^2$. 
The quasi--degenerate spectrum (QD) occurs when the common 
neutrino mass scale $m_0$ 
is much larger than the mass splittings, 
i.e., $m_0^2 \equiv m_3^2 \simeq m_2^2 
\simeq m_1^2 \gg \Delta m^2_{\odot, \rm A}$.\\

\noindent Mass matrices for Dirac neutrinos are not symmetric. Hence, 
they are parameterized by nine complex variables $a,\ldots,l$:   
\be \label{eq:mnu}
m_\nu = 
\left( 
\bad 
a & b & d \\
e & f & g \\
h & k & l 
\ea 
\right) = U \, m_\nu^{\rm diag} \, V^\dagger ~, 
\ee
where $U,V$ are unitary matrices and $m_\nu^{\rm diag}$ is a diagonal 
mass matrix containing the three mass eigenstates. 
In the basis in which the charged lepton mass matrix is diagonal  
the relevant Lagrangian reads 
\be
{\cal  L} = - \overline{(\nu'_\ell)_L} \,  m_\nu \, 
(\nu'_\ell)_R - \overline{\ell_L} \, 
m_\ell^{\rm diag} \, \ell_R + \frac{g}{\sqrt{2}} W_\mu \, \overline{\ell_L} \, 
\gamma^\mu  (\nu'_\ell)_L + h.c.~,
\ee
where $\ell = e, \mu, \tau$.  
After substituting $(\nu'_\ell)_L = U \, (\nu_\ell)_L$ 
and $(\nu'_\ell)_R = V \, (\nu_\ell)_R$
we identify the PMNS matrix $U$. This matrix is associated with the 
diagonalization of $m_\nu \, m_\nu^ \dagger$, i.e.,  
\be \label{eq:h}
U^\dagger\, m_\nu \, m_\nu^ \dagger \, U 
\equiv U^\dagger \, h \,  U 
= (m_\nu^{\rm diag})^2~.  
\ee
Here we also defined the matrix $h = m_\nu \, m_\nu^ \dagger$. 
The matrix $V$ diagonalizes $m_\nu^\dagger \, m_\nu$ and 
is, in the absence of right--handed currents, unobservable.
In case of Majorana neutrinos, we would have $V = U^\ast$. \\

\noindent One of the most interesting aspects of neutrino physics is 
the possibility of leptonic $CP$ violation. 
Whether a given neutrino mass matrix implies $CP$ violation 
in oscillation experiments (note that due to the Dirac nature assumed in 
this work there are no so--called Majorana phases \cite{Majpha}) 
can for instance be determined from the characteristic matrix $h$. 
In terms of $h$, one can write the Jarlskog invariant $J_{CP}$ as \cite{JCP} 
\bea \label{eq:JCP}
J_{CP} = -\frac{\D {\rm Im} (h_{12} \, h_{23} \, h_{31} )}
{\D \Delta m^2_{21} \, \Delta m^2_{31} \, \Delta m^2_{32}} \\[0.3cm]
= \frac{\D 1}{\D 8} \, \sin 2 \theta_{12} \,   \sin 2 \theta_{23} \,  
 \sin 2 \theta_{13} \,  \cos \theta_{13} \, \sin \delta~. 
\eea
The second line in 
Eq.\ (\ref{eq:JCP}) gives the form of $J_{CP}$ in terms of the 
parametrization Eq.\ (\ref{eq:Upara}). 
Hence, if one of the entries $h_{12}$, $h_{13}$ or $h_{23}$ vanishes, 
there is no $CP$ violation in oscillation experiments.\\

\noindent It is helpful to investigate 
the approximate form of $h$ as implied by current data. 
We will see below that $\lambda \simeq \theta_C \simeq 0.22$ is a 
useful parameter to describe 
the order of magnitude of the entries in $m_\nu$. Since it holds to a good 
precision that $U_{e2} = \sqrt{1/2}~(1 - \lambda)$ and 
$R = D~\lambda^2$ (where $D$ is of order one) \cite{ichPRD}, we can write 
in case of normal hierarchy $m_3^2 \simeq \dma$, $m_2^2 \simeq 
\dma \, D~\lambda^2$ and in case of inverted hierarchy 
$m_1^2 \simeq \dma$, $m_2^2 \simeq \dma \, (1 + D~\lambda^2)$. 
Ignoring the smallest mass state and the $CP$ phase and setting for instance 
$U_{e3}=A~\lambda^2$ and $U_{\mu 3} = \sqrt{1/2}~(1 - B \, \lambda^2)$, 
where $A,B$ are real parameters of order one, 
we have in case of a normal hierarchy: 
\be \label{eq:NHapp}
h \sim \dma \, 
\left( 
\bad
\lambda^2 & \lambda^2 & \lambda^2 \\
\cdot & 1 & 1\\
\cdot & \cdot & 1 
\ea 
\right)~,
\ee
and for the inverted hierarchy:  
\be\label{eq:IHapp}
h \sim \dma \, 
\left( 
\bad
1 & \lambda^2 & \lambda^2 \\
\cdot & 1 & 1\\
\cdot & \cdot & 1 
\ea 
\right)~,
\ee
where the entries are to be understood as order--of--magnitude--wise. 
These two simple matrices are a very helpful guiding line in the search for 
promising candidates. 

\noindent Turning to the quasi--degenerate scheme, 
and in case of a normal ordering, we can write 
$m_3^2 = m_0^2$, $m_2^2 = m_0^2~(1 - \eta (1 - D \, \lambda^2))$ and 
$m_1^2= m_0^2~(1 - \eta)$, where $\eta \equiv \dma/m_0^2$. Then, with the 
above definition of the mixing parameters, we find 
\be\label{eq:QDapp}
h \sim m_0^2 \, 
\left( 
\bad
1  + \eta & \eta \, \lambda^2 & \eta \, \lambda^2 \\
\cdot & 1 + \eta & \eta + \lambda^2 \\
\cdot & \cdot & 1 + \eta + \lambda^2
\ea 
\right)~. 
\ee 
A very similar expression is found for the inverted mass ordering. 
Note that the entries of order ``1'' in the above matrix are 
in our approximation exactly 1. 
This means, that the parameters in $m_\nu$ are such that 
the leading terms in the diagonal entries of $h$ are identical.\\

\noindent 
We end our general considerations about Dirac mass matrices with the 
prospects of distinguishing Dirac from Majorana neutrinos. 
Unfortunately, proving that neutrinos are Dirac particles is much 
more difficult than proving them to be Majorana particles. 
The only known process sensitive to the lepton number violation 
characteristic for light Majorana neutrinos is neutrinoless 
double beta decay \cite{APS0vbb}. This process is triggered by the 
so--called effective mass \meff, the $ee$ entry of the Majorana mass matrix 
in the charged lepton mass basis (for details, see, e,g, the recent reviews 
\cite{APS0vbb,STP0vbb}).   

\noindent Suppose first that (except for the oscillation results) there is 
no signal indicating neutrino mass from direct 
laboratory searches or cosmological observations, i.e., we do not 
know the exact neutrino mass scale. 
If the mass ordering is shown to be inverted, we know that 
--- if neutrinos are Majorana particles --- there is a lower 
limit on the effective mass \cite{APS0vbb,STP0vbb} of 
$\meff > \sqrt{\dma} \, 
\cos^2 \theta_{13} \, \cos 2 \theta_{12} \simeq 0.01$ eV. 
Hence, pushing the limit on the effective mass lower than this value 
would show that neutrinos are Dirac particles. 
In case of the normal ordering however, the effective mass can vanish 
even though neutrinos are Majorana particles. 
This would indicate that the $ee$ entry of the neutrino (Majorana) mass 
matrix in the charged lepton mass basis is zero, 
which would be a very interesting observation by itself. 
Nevertheless, we have no handle on determining the nature of the neutrino. 
A possible way out of this dilemma would be to look for analogous processes of 
neutrinoless double beta decay. However, the associated branching ratios 
or cross sections are many orders of magnitude below current and 
future experimental bounds \cite{DL2}.

\noindent If we have however a definite value of the neutrino mass scale, 
be it from cosmology or direct laboratory searches, and do not find a 
corresponding signal in neutrinoless double beta decay experiments, 
we could conclude that neutrinos are Dirac particles. To be more 
quantitative, if the neutrino mass scale is given by $m_0$, 
the effective mass should be larger than roughly $m_0 \cos 2 \theta_{12}$ 
if neutrinos are Majorana particles.

\section{\label{sec:min}Minimal Dirac Mass Matrices}

\noindent In the remainder of the paper we shall consider minimal 
Dirac mass matrices, i.e.\ matrices possessing a certain 
number of zero entries. In the numerical search for successful matrices, 
we required the neutrino mixing observables 
$\sin^2 \theta_{12, 23, 13}$ and $R= \dms/\dma$ to lie 
within their $2\sigma$ ranges given in 
Eq.\ (\ref{eq:valle}). The non--vanishing 
entries are varied within two orders of magnitude, i.e., we allow for 
up to a factor of 100 between them. 
We checked that the found textures are stable with respect to 
small perturbations, in the sense that the results stay unchanged as 
long as the ``zero entries`` are one order of magnitude smaller than the 
non--zero entries.

\subsection{\label{sec:gen_min}General Considerations}

\noindent It is fairly easy to see that the possible 
Dirac nature of neutrinos allows for at least as many 
zeros as in case of Majorana neutrinos. 
Consider in Eq.\ (\ref{eq:mnu}) the case $V = \mathbbm{1}$. Then we have 
$m_\nu = U \, m_\nu^{\rm diag}$ and therefore 
\be \label{eq:V=1}
m_\nu = 
\left( 
\bad
U_{e1} \, m_1 & U_{e2} \, m_2 & U_{e3} \, m_3 \\
U_{\mu 1} \, m_1 & U_{\mu 2} \, m_2 & U_{\mu 3} \, m_3 \\
U_{\tau 1} \, m_1 & U_{\tau 2} \, m_2 & U_{\tau 3} \, m_3 
\ea
\right)~. 
\ee
In the case of an extreme normal (inverted) 
hierarchy we can set $m_1~(m_3) = 0$ 
and thus have three zeros in the first (third) column. Further setting 
$U_{e3}$ to zero, which is compatible with current data, adds another 
zero entry to $m_\nu$ for the normal hierarchical case. We will indeed 
encounter successful texture zero matrices with such a 
structure in the following Sections.\\

\noindent 
One can group Dirac neutrino mass matrices 
into classes which lead to the same matrix $h$. Matrices contained in these
classes can be transformed into one another by a permutation of the
right--handed neutrino fields, which corresponds to an exchange of columns in 
$m_\nu$, i.e., 
\begin{equation}
m_{\nu} \rightarrow  m_{\nu}^{\prime}= m_{\nu} P_{i}~.  
\end{equation}
This leaves the rest of the Lagrangian invariant and leads to 
the same matrix $h$ because 
\[
h=m_{\nu} m_{\nu}^{\dagger} \rightarrow 
h^{\prime}= m_{\nu} ^{\prime} m_{\nu}^{\prime \, \dagger} 
= m_{\nu} P_{i} P_{i}^{\dagger} m_{\nu}^{\dagger} = h ~. 
\]
The orthogonal permutation matrices $P_{i}$ are:
\be \label{eq:P}
\bad P_{1} = \left(\begin{array}{ccc}
1&0&0\\
0&1&0\\
0&0&1
\end{array}\right) ~, ~
P_{2}=\left(\begin{array}{ccc}
0&1&0\\
1&0&0\\
0&0&1
\end{array}\right) ~, ~
P_{3}=\left(\begin{array}{ccc}
0&0&1\\
0&1&0\\
1&0&0
\end{array} \right) ~,~ \\ \\
P_{4}=\left(\begin{array}{ccc}
1&0&0\\
0&0&1\\
0&1&0
\end{array} \right) ~,~
P_{5}=\left(\begin{array}{ccc}
0&0&1\\
1&0&0\\
0&1&0
\end{array} \right) ~, ~
P_{6}=\left(\begin{array}{ccc}
0&1&0\\
0&0&1\\
1&0&0
\end{array} \right)~.
\ea 
\ee
Note that
we cannot put two matrices which are transformed into one another by a
transformation of the left--handed lepton fields in one class, since we
have already fixed the basis in which the mass matrix of the
charged leptons is diagonal and therefore such a transformation would
not leave the rest of the Lagrangian invariant.\\

\noindent Finally, before proceeding with the search for successful 
texture zero mass matrices, it is worth 
commenting on the RG running of a Dirac neutrino 
mass matrix. The non--diagonal entries in the beta function for the 
Yukawa couplings are (in the charged lepton mass basis) proportional to 
\cite{MLD} $m_\nu \, m_\nu^\dagger /v^2$, 
where $v$ is the weak scale. In principle, 
this term could destroy the texture zeros under consideration. However, 
the smallness of the Yukawa couplings, which are in the SM 
at most of order $10^{-11}$ and in the MSSM of order 
$10^{-11} \, (1 + \tan^2 \beta)$, renders the corrections negligible. 
Thus, ignoring possible SUSY threshold corrections, 
the stability of the texture zeros, which will be 
discussed in the next Section, is to a good precision guaranteed.

\subsection{\label{sec:5}Five Texture Zeros}

When the current neutrino data is taken into account, the maximal 
number of zeros in a Dirac mass matrix turns out to be 5. 
There are in total 126 possibilities to put 
5 zeros in $m_\nu$, 18 of which  
allow for a successful explanation of the neutrino data. 
Their common feature is that their determinant vanishes and therefore they  
possess one vanishing mass eigenvalue ($m_{1}=0$
for NH and $m_{3}=0$ for IH). Furthermore, there is no 
$CP$ violation.

\noindent These 18 matrices fall into three classes which we denote 
$A$, $B$ and $\tilde{B}$. Matrices belonging to class $A$ predict the 
inverted hierarchy and $U_{e 3}=0$ 
whereas those belonging to $B$ and $\tilde{B}$
can accommodate both hierarchies and lead in general to non--vanishing
$U_{e3}$. 
 
\noindent Class $A$ is represented by the matrix:
\begin{equation}\label{eq:A}
A = 
\left( 
\bad 
a & b & 0  \\
0 & d & 0 \\
0 & e & 0 
\ea
\right)~, 
\end{equation}
which can be transformed into five other matrices with one of the
permutation matrices $P_{i}$ of Eq.\ (\ref{eq:P}):
\[
A \, P_2 = \left( \begin{array}{ccc} 
    b&a&0\\
    d&0&0\\
    e&0&0
\end{array} \right) ~, ~
A \, P_{3} = \left( \begin{array}{ccc} 
    0&b&a\\
    0&d&0\\
    0&e&0
\end{array} \right) ~, ~
A \, P_{4} = \left( \begin{array}{ccc} 
    a&0&b\\
    0&0&d\\
    0&0&e
\end{array} \right) ~, ~
\] 
\[
A \, P_{5} = \left( \begin{array}{ccc} 
    b&0&a\\
    d&0&0\\
    e&0&0
\end{array} \right) ~, ~
A \, P_{6} = \left( \begin{array}{ccc} 
    0&a&b\\
    0&0&d\\
    0&0&e
\end{array} \right)~.
\]
Next, we show that class $A$ works only in case of an inverted 
hierarchy with the additional constraint of $U_{e3}=0$. 
Note that the matrices of class $A$ predict for the resulting $h$ a vanishing 
(23)--subdeterminant, independent of the choice of the variables $a,b,d,e$. 
Using the definition $h = U \, (m_\nu^{\rm diag})^2 \, 
U^\dagger$ and requiring that 
the (23)--subdeterminant vanishes, leads to the condition  
$( m_3^2 \, c_{13}^2 \left(m_2^2 \, c_{12}^2 
+ m_1^2 \, s_{12}^2 \right) + m_1^2 \, m_2^2 \, s_{13}^2) =0$. 
For a normal hierarchy (i.e., $m_1=0$) this gives 
$m_{2}^{2} m_{3}^{2} c_{12}^{2} c_{13}^{2} = 0$, 
which certainly cannot accommodate the data, since it implies $U_{e1}=0$. 
In the case of an inverted hierarchy (i.e., $m_{3}=0$) 
we end up with the equation $m_{1}^{2} m_{2}^{2} s_{13}^{2} = 0$, 
which implies $\theta_{13}=0$ and leaves the other 
mixing angles unconstrained. 

\noindent In general, the formulae for the neutrino mixing observables 
in terms of $a,\ldots,l$ are quite complicated, 
but in case of 5 zero entries one can obtain simple expressions. 
Taking $a,b,d,e$ real and positive we find 
\bea \label{eq:Aresex}
\tan^2 \theta_{23} = \frac{\D e^2}{\D d^2} ~~,~~ \sin^2 \theta_{12} = 
\frac{1}{2} \left(1 + \frac{\D a^2 + b^2 - d^2 - e^2}{\D w} \right)~~,~~ 
\\[0.3cm]
R = 
\frac{\dms}{\dma} = \frac{\D w}{\frac{1}{2} \D (a^2 + b^2 + d^2 + e^2 - w)}
~~,~~ \\[0.4cm]
\mbox{ where } 
w = \sqrt{(a^2 + b^2 + d^2 + e^2)^2 - 4 a^2 (d^2 + e^2)}~~.
\eea

\noindent Typically, $a$, $e$ and $d$ have 
(in units of $\sqrt{\dma}$) absolute values of order one and 
$b$ is of order $\lambda^2$. With these estimates, one obtains 
the approximate form of $h$ from Eq.\ (\ref{eq:IHapp}). 
Note that $\sin^2 \theta_{12}$ can give the value 
$\sin^2 \theta_{12} \simeq 0.3$ quite naturally. 
However, cancellations of order one parameters lead to the smallness of $R$
which induces a certain fine--tuning. This will be the case for most 
of the solutions which predict the inverted hierarchy. There is typically 
a large (up to two orders of magnitude) gap between the entries of 
$m_\nu$ and this implies that the inverted hierarchy will 
--- in a bottom--up approach --- always be associated with some tuning.\\

\noindent The 12 other matrices able to reproduce the neutrino data 
can be divided into two very similar classes, which we denote 
$B$ and $\tilde{B}$. They generate --- independent of $a,b,d,e$ --- 
a vanishing $h_{12}$ and $h_{13}$, respectively. 
Their representants are
\begin{equation}
B=\left(\begin{array}{ccc}
0 & a & 0 \\
b & 0 & 0 \\
d & e & 0 
\end{array} \right)
\;\; \mbox{and} \;\;
\tilde{B} = 
\left( 
\begin{array}{ccc}
0 & a & 0  \\
b & d & 0 \\
e & 0 & 0 
\end{array}
\right)~.
\end{equation}
Due to the vanishing of $h_{12}$ ($h_{13}$) in class $B$
($\tilde{B}$) $J_{CP}$ is also zero, i.e.\ there is no $CP$ violation. 
From these conditions one can obtain an 
interesting correlation between the neutrino mixing observables. 
Setting the 12 element of $h =  U \, (m_\nu^{\rm diag})^2 \, 
U^\dagger$ to zero gives\footnote{We have assumed $c_{13} \neq 0$.} 
\be \label{eq:12zero}
|U_{e3}| = \frac{1}{2} \frac{R \, \sin 2 \theta_{12} \, 
\cot \theta_{23} }{1 \mp R \, \sin^2 \theta_{12}} 
\simeq \frac{1}{2} R \, \sin 2 \theta_{12} \, 
\cot \theta_{23}~,
\ee
where the $-$ sign is for the normal and the $+$ for the 
inverted mass ordering. 
The smallness of $R$ implies that the condition $h_{12}=0$ gives to a good 
precision the same correlation for both mass orderings. 
A similar relation can be deduced from $h_{13}=0$ for class $\tilde{B}$: 
\be \label{eq:13zero}
|U_{e3}| = \frac{1}{2} \frac{R \, \sin 2 \theta_{12} \, 
\tan \theta_{23} }{1 \mp R \, \sin^2 \theta_{12}} 
\simeq \frac{1}{2} R \, \sin 2 \theta_{12} \, 
\tan \theta_{23}~. 
\ee
The (close--to--)maximal value of $\theta_{23}$ indicates that 
both, $h_{12}=0$ and $h_{13}=0$, give approximately the same result for 
$U_{e3}$. 
Moreover, the above relations Eqs.\ (\ref{eq:12zero},\ref{eq:13zero}) are 
valid also for a non--vanishing smallest neutrino mass. 
Consequently, we can apply 
them also in later Sections of this work, when $h_{12}$ or $h_{13}$ 
vanish for matrices with a non--vanishing determinant. 
We remark here that one can show that no other element of 
$h$ except for $h_{12}$ and $h_{13}$ 
can vanish without leading to conflicts with the neutrino data. 
We will comment on unsuccessful matrices in Section \ref{sec:dontwork}. 

\noindent Inserting the best--fit values $R= 0.036$,
$\sin^2 \theta_{12}=0.3$ and $\sin^2 \theta_{23}= 0.5$, gives for both 
$h_{12}=0$ and $h_{13}=0$ as well as for both mass orderings that 
$|U_{e3}| \simeq 0.017$ 
which is about two orders of magnitude below the current 2$\sigma$ upper 
bound for $|U_{e3}|^2$. 
Its value increases (decreases) with increasing 
(decreasing) $\sin^2 \theta_{23}$ for a zero 13 entry in 
$h$ (and vice versa for zero $h_{12}$). 
We plot in Fig.\ \ref{fig:fig1} the correlation between 
$\sin^2 \theta_{23}$ and $|U_{e3}|^2$ for class 
$\tilde{B}$ in case of the normal hierarchy. 
Fig.\ \ref{fig:fig1a} displays the same for matrix 
$B$, showing the slightly different behavior for these two 
observables as discussed above and implied by Eqs.\ 
(\ref{eq:12zero},\ref{eq:13zero}). 
The respective plots for the inverted hierarchy look identical, as 
expected from the approximate formulae discussed in this Section. 
As can be seen, $|U_{e3}|^2$ stays between $10^{-4}$ and 
$10^{-3}$; in particular, 
for the best--fit value of $\sin^2 \theta_{23} = 0.5$ we have 
$10^{-4} \ls |U_{e3}^2| \ls 5 \cdot 10^{-3}$.

\noindent Now we investigate the dependence of the observables
on the variables $a,b,d,e$ with $a,b,d,e>0$ in class
$\tilde{B}$. Similar calculations for the class $B$ reveal similar
results and are therefore omitted. 
In case of normal hierarchy, one finds: 
\bea \label{eq:BresNHex}
\tan^2 \theta_{23} = 
\frac{\D (a^2 + b^2 + d^2 - e^2 + w)^2}{\D 4 \, b^2 \, e^2}~~,~~
U_{e1}^2 = \frac{\D d^2 \, e^2}{\D d^2 \, e^2 + a^2 (b^2 + e^2)}~~,~~
\\[0.3cm]
U_{e3} = \frac{\D a (a^2 - b^2 + d^2 - e^2 + w)}
{\D \left( 4 \, b^2 \, d^2 \, e^2 + a^2 (a^2 - b^2 + d^2 - e^2 + w)^2 
+ d^2 (a^2 + b^2 + d^2 - e^2 + w)^2 \right)^{1/2}}~~,~~\\[0.3cm]
R = \frac{\frac{1}{2}  \D 
(a^2 + b^2 + d^2 + e^2 - w)}{\frac{1}{2} \D 
(a^2 + b^2 + d^2 + e^2 + w)} ~~,~~\\[0.3cm]
\mbox{ where } 
w = \sqrt{(a^2  + b^2  + d^2  + e^2 )^2  - 4 (d^2  e^2  + a^2  (b^2  + e^2 ))}
~~.
\eea

\noindent Typically, the normal hierarchy is achieved for parameter values 
(again in units of $\sqrt{\dma}$) of order one for $b,e$ and 
of order $\lambda$ for $a$ and $d$. Inserting those values in $m_\nu$ yields 
the approximate form of $h$ 
from Eq.\ (\ref{eq:NHapp}), where however the 13 entry vanishes. 
With the order of magnitude of $a,b,d,e$ we can obtain from 
Eq.\ (\ref{eq:BresNHex}) the following approximate expressions: 
\bea
\tan^2 \theta_{23} \simeq \frac{\D b^2 }{\D e^2}~~,~~
\sin^2 \theta_{12} \simeq \frac{\D \tilde{a}^2 (b^2+e^2)} 
{\D \tilde{a}^2 b^2 + e^2 (\tilde{a}^2 + \tilde{d}^2)} ~~,~~\\[0.4cm]
U_{e3} \simeq \frac{\D \tilde{a} \tilde{d} \, b}{\D (b^2 + e^2)^{3/2}} 
\, \lambda^2~~,~~
R \simeq \frac{\D \tilde{d}^2 e^2 + \tilde{a}^2 (b^2 + e^2 )} 
{\D (b^2 + e^2)^{2}} \, \lambda^2 ~,
\eea
where we have defined new order one parameters 
$\tilde{a} \equiv a/ \lambda$ and $\tilde{d} \equiv d/ \lambda$. 
As can be seen, $U_{e3}$ and $R$ are small because they are proportional to 
$\lambda^{2}$ and, again, $\sin^2 \theta_{12}$ can be naturally of order one.

\noindent In case of inverted hierarchy, we get: 
\bea \label{eq:BresIHex}
\tan^2 \theta_{23} = \frac{\D e^2}{\D b^2} 
~~,~~ U_{e3} = \frac{\D d e}{\D \sqrt{ a^2 b^2 + e^2 (a^2 + d^2 )} }
~~,~~\\[0.4cm] 
\sin^2 \theta_{12} = \frac{\D 1}{\D 2} \left(1 - 
\frac{\D e^4 - e^2 (a^2 - 2b^2 + d^2) - b^2 (a^2 - b^2 - d^2)}
{\D (b^2 + e^2) w}
\right)
 ~~,~~ \\[0.4cm] 
R = \frac{\D w}{\frac{1}{2} \D 
(a^2 + b^2 + d^2 + e^2 - w)} ~~,~~\\[0.4cm] 
\mbox{ where } 
w = \sqrt{(a^2  + b^2  + d^2  + e^2 )^2  - 4 (d^2  e^2  + a^2  (b^2  + e^2 ))}
~~.
\eea
In order to explain the neutrino data one needs now that $a,b,e$ are 
of order one and $d$ of order $\lambda^2$ (leading approximately to 
Eq.\ (\ref{eq:IHapp}) with a vanishing 13 entry).  
Analyzing Eq.\ (\ref{eq:BresIHex}) shows that $R$ owes its 
smallness to a cancellation of the order one coefficients which is similar 
to the matrices in class $A$ discussed above.

\noindent 
To conclude this Section, both the normal and the inverted hierarchy 
can be accommodated with Dirac neutrino mass matrices containing 5 
zero entries. No observable $CP$ violation is predicted and $U_{e3}$ 
is either exactly zero (then the inverted hierarchy is present) 
or very small, corresponding to $|U_{e3}|^2 \simeq 10^{-4} \ldots 10^{-3}$. 
A non--vanishing value of $|U_{e3}|^2$ above $10^{-3}$ will therefore 
rule out any Dirac mass matrix with 5 zero entries, while a value between 
$10^{-4}$ and $10^{-3}$ excludes class $A$. Class $A$ is also 
ruled out when the neutrino mass spectrum is found to be 
normally ordered.   
The three classes of matrices $A,B$ and $\tilde{B}$ 
are furthermore excluded when either leptonic $CP$ violation is present or 
when a positive signal of neutrino mass in laboratory searches or 
by cosmological observations is found. 
Distinguishing $B$ from $\tilde{B}$ will be very difficult, unless 
$\theta_{23}$ deviates sizably from $\pi/4$. 
We give in Table \ref{tab:tab5yes} the classes of matrices which 
are successful in explaining the neutrino data. Together with the number 
of members in the class we give their phenomenological consequences and 
an order of magnitude estimate of the non--zero entries.

\subsection{\label{sec:4}Four Texture Zeros}

As shown in the previous Subsection, Dirac mass matrices with five 
zero entries 
can accommodate neutrino data, but predict in general 
vanishing $CP$ violation. 
To take this effect into account 
enforces one to 
allow for an additional non--zero entry in $m_\nu$. 
As for the case of 5 zero entries, there are 126 possible matrices, 69 
of which are successful in explaining the data. Those 69 
belong to 12 classes. Among those are 4 
classes with a vanishing determinant.  From those 4 there are 3 classes 
which can accommodate leptonic $CP$ violation. 
Let us start with the $CP$ conserving matrices.\\

\noindent The first class, denoted $C$ and containing 3 members, is 
represented by 
\begin{equation}
C=\left( 
\begin{array}{ccc} 
a & b & d \\
e & 0 & 0 \\
f & 0 & 0 
\end{array}
\right)~.
\end{equation}
The resulting structure of $h$ is analogous to the one of 
class $A$ in the previous Subsection, hence one massless neutrino, 
zero $U_{e3}$ and the inverted hierarchy are predicted. At least 
one of the entries has to be 
suppressed by $\lambda^2$ in order to reproduce the approximate 
texture of Eq.\ (\ref{eq:IHapp}).

\begin{figure}[t]
\begin{center}
\epsfig{file=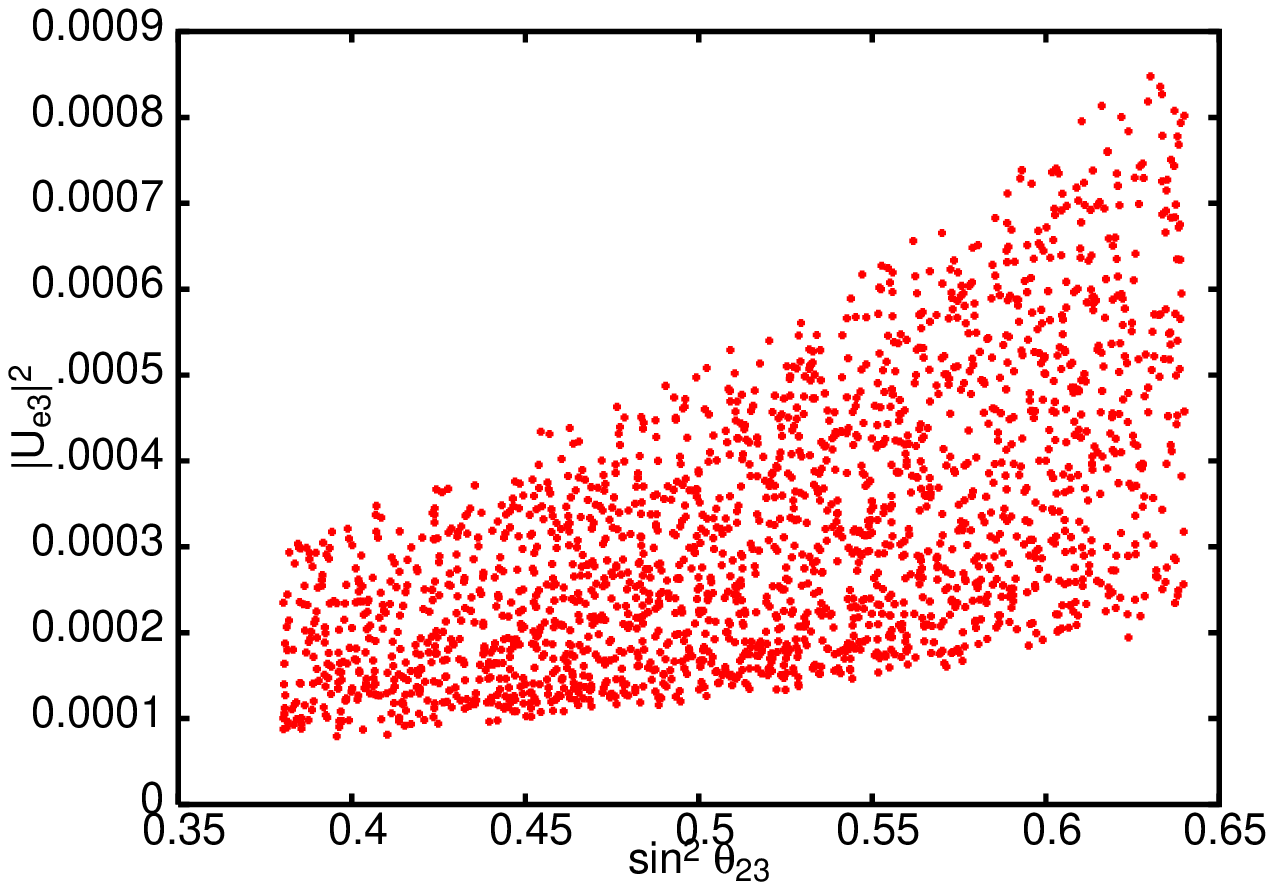,width=14cm,height=9cm}
\caption{\label{fig:fig1}Scatter plot of the correlation 
between $\sin^2 \theta_{23}$ and $|U_{e3}|^2$ for class $\tilde{B}$ in case 
of normal hierarchy. All matrices associated with a zero entry in 
$h_{13}$ generate the same behavior.}
\end{center}
\begin{center}
\epsfig{file=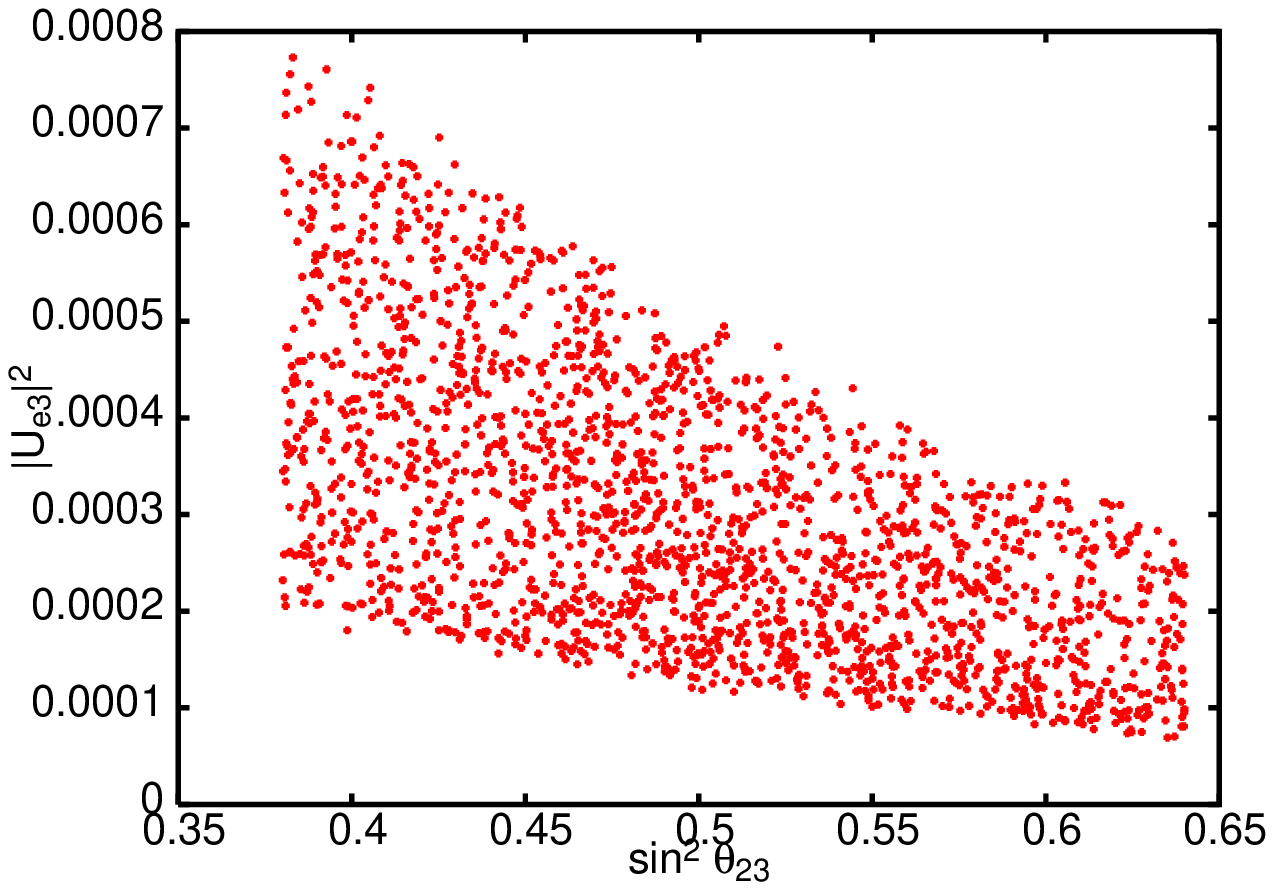,width=14cm,height=9cm}
\caption{\label{fig:fig1a}Same as above for class $B$. 
All matrices associated with a zero entry in 
$h_{12}$ generate the same behavior.}
\end{center}
\end{figure}

\noindent Other classes have a vanishing 12 or 13 entry of $h$ 
and therefore, as argued above, a vanishing Jarlskog invariant.  
The resulting correlations between the observables, as given in 
Eqs.\ (\ref{eq:12zero},\ref{eq:13zero}) and shown in 
Figs.\ \ref{fig:fig1} and \ref{fig:fig1a}, hold again in these cases. 
Consider the following 18 matrices, 
which fall into the three classes $D_{1,2,3}$ and predict $h_{12}=0$: 
\be\label{eq:12D}
D_1 = 
\left( 
\bad 
a & 0 & 0 \\
0 & b & 0 \\
d & e & f 
\ea
\right)~~, ~~
D_2 = 
\left( 
\bad 
a & 0 & 0  \\
0 & b & d \\
e & 0 & f 
\ea
\right)~~,~~
D_3 = 
\left( 
\bad 
a & b & 0  \\
0 & 0 & d \\
e & 0 & f 
\ea
\right)~~.
\ee
From Eq.\ (\ref{eq:12zero}), one has 
$|U_{e3}| \simeq \frac{1}{2} R \, \sin 2 \theta_{12} \, \cot \theta_{23}$, 
whose behavior is displayed in Fig.\ \ref{fig:fig1a}. 
To accommodate a normal hierarchical spectrum, 
the second class $D_2$ has typically $a,b,e$ of order $\lambda$ 
and $d,f$ of order one. 
If $e$ is of order $\lambda^2$, 
$d$ of order $\lambda$ and the other parameters of order 1, it turns out 
that the spectrum can become quasi--degenerate. This can be seen from the 
fact that the determinant of the matrix,  
${\rm det}~ D_2 \, D_2^\dagger= a^2 \, b^2 \, f^2 $ is then close 
to the third power of the trace, 
${\rm Tr}~ D_2\, D_2^\dagger = (a^2 + b^2 + d^2 + e^2 + f^2)$. 
Recall that in a bottom--up approach, 
the case of quasi--degenerate neutrinos is 
fine--tuned, since the diagonal elements of $h$ have to be identical. 
The inverted mass ordering is also possible, but as in the classes 
$B$ and $\tilde{B}$ discussed for matrices with 5 zeros, 
this is connected with a large spread of parameters. 
To be precise, the entries $b$ and $e$ of $D_2$ have to be of 
order $\lambda^2$ and $a,d,f$ of order 1. 

\noindent The following three classes $\tilde{D} _{1,2,3}$, 
giving in total 18 matrices, generate a zero in $h_{13}$: 
\be \label{eq:13E}
\tilde{D}_1 = 
\left( 
\bad 
a & 0 & 0 \\
b & d & e \\
0 & f & 0 
\ea
\right)~~, ~~
\tilde{D}_2
\left( 
\bad 
a & 0 & 0  \\
b & d & 0 \\
0 & e & f 
\ea
\right)~~,~~
\tilde{D}_3 = 
\left( 
\bad 
a & b & 0  \\
d & 0 & e \\
0 & 0 & f  
\ea
\right)~~.  
\ee
Thus, for them Eq.\ (\ref{eq:13zero}) holds and the behavior of $U_{e3}$ and 
$\sin^2 \theta_{23}$ can be seen in Fig.\ \ref{fig:fig1}. Similar 
statements about the neutrino mass scheme as for the classes $D_{1,2,3}$ 
can be made.\\

\noindent We furthermore find two classes denoted $E$ and $\tilde{E}$, 
which contain six matrices each.  Representants look like:
\be \label{eq:E}
E= 
\left( 
\bad 
a & b & 0 \\
d & 0 & 0 \\
e & 0 & f 
\ea
\right)~~\mbox{ and }~~
\tilde{E}= 
\left( 
\bad 
a & 0 & b  \\
d & e & 0 \\
f & 0 & 0 
\ea
\right)~~.
\ee
Their common feature is a non--vanishing determinant and 
$J_{CP}=0$. From the matrices one finds that 
\be \label{eq:classE}
\frac{h_{12}}{h_{13} } = \frac{h_{22}}{h_{23} } 
\ee
for class $E$ while for class $\tilde{E}$ it holds 
\be \label{eq:classEt}
\frac{h_{12}}{h_{13}} = \frac{h_{32}}{h_{33}}~.
\ee
Inserting Eq.\ (\ref{eq:classE}) in the definition of $J_{CP}$ 
from Eq.\ (\ref{eq:JCP}) gives for class $E$ that  
\be 
J_{CP} \propto {\rm Im} (h_{12} \, h_{23} \, h_{31} )= {\rm Im}
(h_{22} \, h_{13} \, h_{31}) = 0~,
\ee
since $h$ is a hermitian matrix. A similar equation can be written down for 
class $\tilde{E}$. 
This explains why there is no 
$CP$ violation for the classes $E$ and $\tilde{E}$. 

\noindent The conditions resulting from Eqs.\ (\ref{eq:classE}) 
and (\ref{eq:classEt}) can be used to obtain an interesting 
correlation between the neutrino mixing observables. 
Starting with class $E$, the relation Eq.\ (\ref{eq:classE}) leads 
for the normal mass ordering to: 
\bea \label{eq:EcorNH} 
|U_{e3}| = 
\frac{\D \dms (\dma + m_1^2) \sin 2 \theta_{12} \, \tan \theta_{23}}
{\D \dms \dma + (2\,  \dma - \dms) m_1^2 + \dms (\dma + m_1^2) \, 
\cos 2 \theta_{12}}\\[0.3cm] \\
\simeq 
\left\{ 
\baz
\tan \theta_{23} \, \tan \theta_{12} & \mbox{ for }  
m_1^2 = 0 \\[0.2cm]
\frac{\D 1}{\D 2} R \, \sin 2 \theta_{12} \, \tan \theta_{23} & \mbox{ for }  
m_1^2 \gg \Delta m^2_{\odot, A} 
\ea 
\right. ~,
\eea
where we gave the limits for the normal hierarchical 
and the quasi--degenerate spectrum. 
Note that the limit for  
$m_1^2 \gg \Delta m^2_{\odot, \rm A}$, which corresponds to a 
quasi--degenerate spectrum, is just the correlation 
of the parameters Eq.\ (\ref{eq:13zero})
for a vanishing 13 element of $h$. 
The fact that the value of $U_{e3}$ in case of 
zero $m_1$ is too large indicates that the matrices of class $E$ lead to a 
lower limit on the neutrino mass if the spectrum is normally ordered. 
Indeed, one can solve Eq.\ (\ref{eq:EcorNH}) for 
$m_1^2/\dms$, which gives: 
\bea \label{eq:a1}
\frac{\D m_1^2}{\dms} = 
\frac{\D \sin 2 \theta_{12} \, \sin \theta_{23} - 2 \, \cos^2 \theta_{12} \, 
\cos \theta_{23} \, |U_{e3}| }
{\D 2 (1 - R \, \sin^2 \theta_{12}) \, \cos \theta_{23} \, |U_{e3}| - 
R \, \sin \theta_{23} \,  \sin 2 \theta_{12} } \\[0.3cm] \\
\simeq 
\left\{ 
\bad 
-\frac{\D 1}{\D R} & \mbox{ for } U_{e3} = 0 \\[0.2cm] 
\cos \theta_{12} 
\left( 
\frac{\D \sin \theta_{12} \, \tan \theta_{23}}{\D |U_{e3}| } 
- \cos \theta_{12} 
\right) 
& \mbox{ for } |U_{e3}| \gg R 
\ea 
\right.~. 
\eea
Consequently,  $m_1^2/\dms$ is minimized for the largest possible $|U_{e3}|$, 
and is then given by $m_1^2 \gs \frac{1}{2} \, 
\dms \, \sin 2 \theta_{12} \, \tan \theta_{23}/|U_{e3}|$. Putting in numbers 
gives $m_1 \gs 0.01$ eV. The mass spectrum can become degenerate when the 
denominator of Eq.\ (\ref{eq:a1}) goes to zero, which occurs when 
$U_{e3}$ takes the value given in Eq.\ (\ref{eq:13zero}). This value 
is about 0.02. Lower values 
of $|U_{e3}|$ render $m_1^2/\dms$ negative and are not possible. 
Hence, $|U_{e3}| \simeq \frac{1}{2} R \, \sin 2 \theta_{12} \, 
\tan \theta_{23}$ is the lowest possible value for class $E$ and the 
normal mass ordering.

\noindent In case of an inverted mass ordering Eq.\ (\ref{eq:classE}) leads
to 
\bea \label{eq:EcorIH} 
|U_{e3}| = 
\frac{\frac{1}{2} 
\D \dms m_3^2 \, \sin 2 \theta_{12} \, \tan \theta_{23}}
{\D \dma (m_3^2 + \dma) + \dms (\dma - m_3^2 \, \sin^2 \theta_{12} )}
\\[0.3cm]\\
\simeq 
\left\{ 
\bad 
0 & \mbox{ for } & 
m_3^2 = 0 \\[0.2cm]
\frac{\D 1}{\D 2} R \, \sin 2 \theta_{12} \, \tan \theta_{23} & \mbox{ for } & 
m_3^2 \gg \Delta m^2_{\odot, A} 
\ea 
\right. ~. 
\eea
In this case there is no lower limit on the 
smallest mass state because $|U_{e3}|$ 
goes to zero when $m_3$ goes to zero. 
One can again solve Eq.\ (\ref{eq:EcorIH}) for the ratio of 
the smallest neutrino mass $m_3$ divided by the solar mass squared difference, 
which gives  
\bea \label{eq:a2} 
\frac{\D m_3^2}{\D \dms } = - 
\frac{\D 2 (1 - R) \cos \theta_{23} \, |U_{e3}| } 
{\D R \left( 2 (1 + R \, \sin^2 \theta_{12} ) 
\cos \theta_{23} \, |U_{e3}| - R \, \sin \theta_{23} \, 
\sin 2 \theta_{12} \right) } \\[0.3cm]\\
\simeq 
\left\{
\bad 
0 & \mbox{ for } U_{e3} = 0 \\[0.2cm]
-\frac{\D 1}{\D R} & \mbox{ for } U_{e3} \gg R
\ea 
\right. ~. 
\eea
The denominator vanishes when $U_{e3}$ fulfills Eq.\ (\ref{eq:13zero}), 
thereby corresponding to quasi--degenerate neutrinos. The expression 
for $m_3^2 /\dms$ becomes negative when $U_{e3}$ is larger 
than roughly 0.02. 
Hence, $|U_{e3}| \simeq \frac{1}{2} R \, \sin 2 \theta_{12} \, 
\tan \theta_{23}$ is the largest possible value for class $E$ and the 
inverted mass ordering. 

\begin{figure}[t]
\begin{center}
\epsfig{file=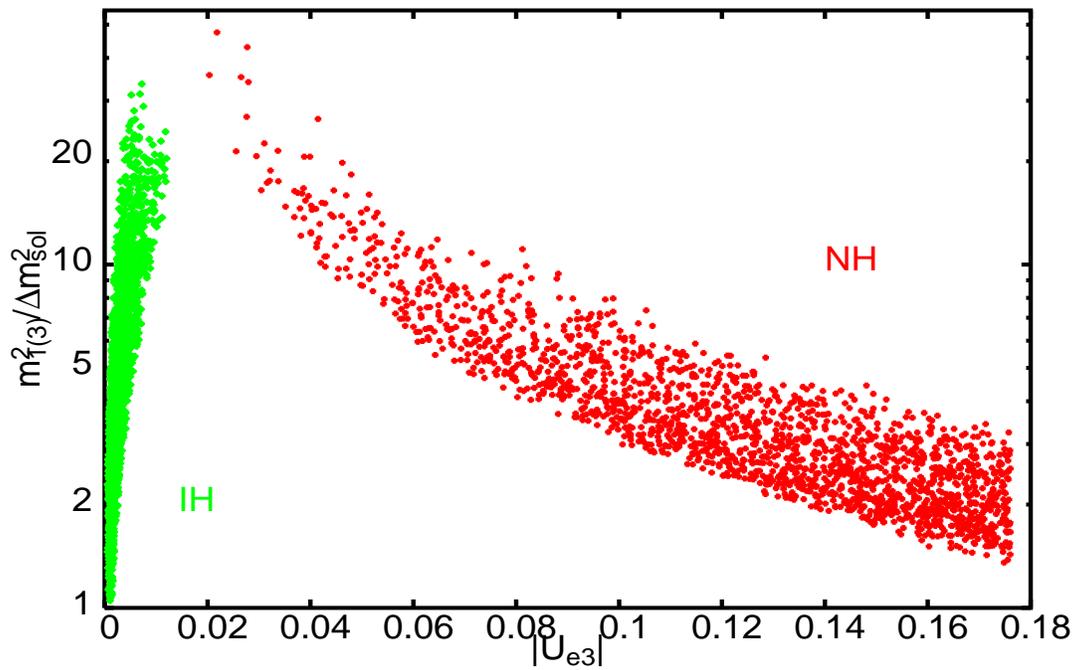,width=14cm,height=9cm}
\caption{\label{fig:fig4}Scatter plot of $|U_{e3}|$ against  
the ratio of the smallest neutrino mass squared divided by \dms{} 
for class $E$, cf.\ Eqs.\ (\ref{eq:a1},\ref{eq:a2}).}
\end{center}
\end{figure}

\noindent We plot in Fig.\ \ref{fig:fig4} for both mass orderings 
and class $E$ a scatter plot of $|U_{e3}|$ against  
the ratio of the smallest neutrino mass squared divided by \dms. 
The lower limit of $|U_{e3}|$, 
roughly 0.02 for the normal mass ordering, occurs for large $m_1^2/\dms$, 
i.e., for quasi--degenerate neutrinos. It is given, as discussed above, 
by $R/2 \, \sin 2 \theta_{12} \, \tan \theta_{23} $. The qualitative 
discussions given above are nicely confirmed by the plot. 

\noindent For the matrices belonging to class 
$\tilde{E}$ we have the relation 
$h_{12}/h_{13} - 
h_{32}/h_{33} = 0$, which leads 
to conditions for the observables very similar to 
Eqs.\ (\ref{eq:EcorNH},\ref{eq:EcorIH}). They are obtained from those 
equations by replacing $\tan \theta_{23}$ with $\cot \theta_{23}$.\\

\noindent 
The remaining classes of 4 zero mass matrices predict non--zero $U_{e3}$ 
and non--vanishing $CP$ violation, which --- from the phenomenological 
point of view --- are perhaps the most interesting ones. 
There are three classes, all of which have a vanishing determinant, 
i.e., one neutrino is massless: 
\be \label{eq:F}
F_1 = 
\left( 
\bad 
a & 0 & 0 \\
b & d & 0 \\
e & f & 0 
\ea
\right)~~, ~~
F_2 = 
\left( 
\bad 
a & b & 0  \\
d & 0 & 0 \\
e & f & 0 
\ea
\right)~~,~~
F_3 = 
\left( 
\bad 
a & b & 0  \\
d & e & 0 \\
f & 0 & 0  
\ea
\right)~~.  
\ee
Unfortunately, there is in general no interesting correlation 
between the parameters and all observables, including the $CP$ phase, 
can take any value inside their allowed regimes. 
Again, the reader can easily obtain the 
rough order of magnitude of the entries in $F_{1,2,3}$ 
in order to reproduce the approximate textures given in 
Eqs.\ (\ref{eq:NHapp},\ref{eq:IHapp}). 
However, in contrast to the case of 5 zero entries, the position of the 
large (i.e., order 1) and small (i.e., order $\lambda$ or $\lambda^2$) terms 
is not unique and there is some freedom. Examples are given in Tables 
\ref{tab:tab4yes1} and \ref{tab:tab4yes2}, together with the implied 
phenomenology and the number of matrices in a given class. 

\noindent We encounter here the simple example given in the beginning 
of Section \ref{sec:gen_min}. 
If the parameters in class $F_1$ are such that 
$F_1^\dagger \, F_1 $ is diagonal, we have $V=\mathbbm{1}$ and 
can accommodate the normal hierarchy together with $U_{e3}=0$, see the 
discussion after Eq.\ (\ref{eq:V=1}). This would require in Eq.\ (\ref{eq:F})  
that $d \, b^\ast + f \, e^\ast = 0$.\\

\noindent 
We can rule out class $C$ with a non--vanishing value of $U_{e3}$ or when 
the mass ordering is normal or when 
laboratory searches or cosmological observations find a signal corresponding 
to a non--zero neutrino mass. 
Classes $D_{1,2,3}$ and $\tilde{D}_{1,2,3}$ 
(which are difficult to distinguish) are ruled out 
for a value of $U_{e3}$ larger or smaller 
than $\simeq R/2 \, \sin 2 \theta_{12}$. They can in principle 
be distinguished from the 5 zero classes $B$ and $\tilde{B}$ because 
they allow for a non--vanishing smallest neutrino mass. 
Classes $E$ and $\tilde{E}$ are difficult to distinguish and are 
excluded when the mass ordering is normal 
(inverted) and $U_{e3}$ is smaller (larger) 
than $R/2 \, \sin 2 \theta_{12}$. 
All the classes $C$, $D_{1,2,3}$, $\tilde{D}_{1,2,3}$, $E$ and $\tilde{E}$ 
are ruled out if there is leptonic $CP$ violation. 
Finally, classes $F_{1,2,3}$ are ruled out if all three neutrinos are 
found to be massive.

\newpage

\subsection{\label{sec:3}Three and less Texture Zeros}
As we have seen, already for matrices with 4 zeros the results for the 
observables can be somewhat arbitrary (see the discussion for the classes 
$F_{1,2,3}$ after Eq.\ (\ref{eq:F})). 
The major part of the successful matrices with 3 or less   
zeros falls into this category. 
Counting the number of possibilities for three zero entries, 
one arrives at 84 of which only 6 generate one vanishing 
eigenvalue. Three of these can accommodate all data and are
represented by:
\begin{equation}\label{3zeromat}
\left(
\begin{array}{ccc}
  a&b&0\\
  d&e&0\\
  f&g&0\\
\end{array} \right)~. 
\end{equation}
No correlation of observables, except that $m_1$ or $m_3$ vanishes, 
is predicted. This provides a possibility to exclude them. 
Again, if the parameters in this matrix are such that $m_\nu^\dagger \, m_\nu$ 
is diagonal, we have $V=\mathbbm{1}$ and can accommodate both 
hierarchies, compare with the discussion after Eq.\ (\ref{eq:V=1}). 

\noindent 
The remaining 78 matrices without vanishing eigenvalues can be divided 
into sixteen classes. 
Two of those, each containing three matrices, fulfill
the relation $h_{12}=0$ and read: 
\[
\left(\begin{array}{ccc}
  0&0&a\\
  b&d&0\\
  e&f&g\\
\end{array} \right)\;\;\mbox{and} \;\; \left(
\begin{array}{ccc}
  0&a&b\\
  d&0&0\\
  e&f&g\\
\end{array} \right)~.
\]
They lead to the same results as the matrices belonging to classes 
$D_{1,2,3}$ (hence are not distinguishable from the latter), 
in particular they predict $CP$ conservation, 
and have the same correlation between $U_{e3}$ and 
$\sin \theta_{23}$ as given in Eq.\ (\ref{eq:12zero}) and 
Fig.\ \ref{fig:fig1a}. 
There also exist two classes with $h_{13}=0$ whose
representants are:
\[
\left(
\begin{array}{ccc}
  0&0&a\\
  b&d&e\\
  f&g&0\\
\end{array} \right)\;\;\mbox{and} \;\; \left(
\begin{array}{ccc}
  0&a&b\\
  d&e&f\\
  g&0&0\\
\end{array} \right)~. 
\]
They lead to the same results as the matrices belonging to classes 
$\tilde{D}_{1,2,3}$ and consequently are not distinguishable from them and 
only very hard to distinguish from the above two classes. 
The presence of $CP$ violation will exclude the last four classes.  
All the other classes contain six matrices each and are represented by:
\bea \label{eq:G}
G_{1}=\left(\begin{array}{ccc}
   0 & 0 & a  \\
   0 & b & d  \\
   e & f & g
\end{array} \right)~,  ~ 
G_{2}=\left(\begin{array}{ccc}
  0  & 0  & a  \\
  b  & d &  e \\
  0  & f & g
\end{array} \right) ~, ~ 
G_{3}=\left(\begin{array}{ccc}
  0  & a & b  \\
  0  & 0 & d  \\
  e  & f & g
\end{array} \right)~, ~ 
G_{4}=\left(\begin{array}{ccc}
  0  & a & b  \\
  0  & d & e  \\
  f  & 0 & g
\end{array} \right)~, ~ 
\\ [0.5cm]
G_{5}=\left(\begin{array}{ccc}
  0  & a  & b  \\
  d  & 0  & e  \\
  0  & f & g
\end{array} \right) ~, ~ 
G_{6}=\left(\begin{array}{ccc}
   0 & a & b  \\
   d & 0 & e  \\
   f & 0 & g
\end{array} \right) ~, ~ 
G_{7}=\left(\begin{array}{ccc}
  0  &a  & b  \\
  d  & e & f  \\
  0  & 0 & g
\end{array} \right) ~, ~ 
G_{8}=\left(\begin{array}{ccc}
   a & b & d  \\
   0 & 0 & e  \\
   0 & f & g
\end{array} \right) ~, ~ 
\\[0.5cm]
G_{9}=\left(\begin{array}{ccc}
   a & b & d  \\
   0 & e & f  \\
   0 & 0 & g
\end{array} \right)~,~ 
G_{10}=\left(\begin{array}{ccc}
  0  & a & b  \\
  d  & 0 & e  \\
  f  & g & 0
\end{array} \right)~.
\eea
None of these reveal simple correlations among the observables. 

\noindent Similarly, the 36 possible matrices with two texture zeros
and the nine ones with only one zero 
(of course, all of them are allowed) do not show any
relations among the mixing angles, the $CP$--phase and the ratio of the
mass squared differences. 

\subsection{Symmetric Matrices with Texture Zeros}
The discussion so far assumed a general structure for the 
Dirac neutrino mass matrices. In a given model based on some symmetry, 
however, symmetric matrices might arise. 
Extending our discussion to the case of symmetric Dirac mass matrices 
is rather straightforward. 
It is known from the discussion of Ref.\ \cite{2zeros0} 
that Majorana mass matrices, which are necessarily symmetric, can only have 
two independent zero entries. 
It turns out now that if the three zero Dirac mass matrices in the 
classes $G_i$ (see Eq.\ (\ref{eq:G})) are symmetric, they can 
reproduce six of the seven allowed two zero Majorana mass matrices 
from Refs.\ \cite{2zeros0}. In the language of those references, 
the two zero Majorana mass matrix $A_1$ corresponds to the symmetric form 
of matrix $G_1$ from Eq.\ (\ref{eq:G}). 
Analogously, it is easy to see that $G_2, G_3, G_7$ 
from Eq.\ (\ref{eq:G}) correspond to $A_2, B_3, B_4$ 
from Refs.\ \cite{2zeros0}, respectively. 
The matrices belonging to class $G_{10}$ reproduce cases $B_1$ and $B_2$. 
We display this correspondence in Table \ref{tab:FGM}. 
Case $C$ from \cite{2zeros0} would correspond to a 
symmetric mass matrix with two zero entries, whose discussion we omitted 
in this work, since in general no interesting correlation of 
variables arises from this possibility. 
The phenomenological consequences of two zero Majorana mass matrices are 
rather well known \cite{2zeros0} and we have nothing to add here and 
refer to those works for details of their phenomenological implications. 
Nevertheless, we included the approximate formulae out of which the 
phenomenology follows in the Table.

\subsection{\label{sec:dontwork}Unsuccessful Matrices}

We finish this Section with a few comments on matrices which are
unsuccessful in explaining the neutrino data. 

\noindent In case of 5 zeros in $m_{\nu}$, there are three categories
of matrices which do not work. First, there are 90 matrices which lead
--- independently of the choice of the four complex parameters --- to
the vanishing of two mixing angles in $U$. 
In the second category, there are twelve
matrices which generate a zero (12)-- or (13)--subdeterminant in $h$. 
A vanishing of the (12)--subdeterminant leads for the 
normal hierarchy ($m_{1}=0$) to $U_{\tau 1}=0$ and 
for the inverted hierarchy ($m_{3}=0$) to $U_{\tau 3}=0$. Analogously, if 
the (13)--subdeterminant vanishes, one ends up with $U_{\mu 1}=0$ for the 
normal mass ordering and $U_{\mu 3}=0$ for the inverted ordering. All
these conditions are of course not compatible with the data. In the
last category, there are six matrices predicting $h_{23}=0$. 
Phenomenologically $h_{23}=0$ leads to a too large value
for $U_{e 3}$ for the normal as well as the inverted hierarchy. Hence, 108 out
of 126 possible matrices with 5 zeros are unsuccessful in explaining 
the neutrino data.

\noindent For matrices with four zeros we find that about half of the
126 possible matrices, i.e., 57 matrices, do not work. 27 matrices --- 9 of
which have a non--vanishing determinant -- are not successful since
two of the mixing angles are zero. Furthermore, three matrices generate a
vanishing (12)--subdeterminant in $h$ and further three ones a
vanishing (13)--subdeterminant. Apart from these there are 18 matrices
leading to $h_{23}=0$. Finally, there are six matrices fulfilling the
equation $h_{11}/h_{12}=h_{31}/h_{32}$. 
In this case $|U_{e3}|$ turns out to be always bounded from 
below by $\simeq 0.3$ and hence one cannot accommodate the data. 

\noindent In Table \ref{tab:dontwork} we give typical examples of matrices 
with 4 and 5 zero entries which are not able to explain the data.

\noindent Only 9 of the possible 84 three zero textures do not work. Three of
them enforce two mixing angles to be zero and the six other ones
generate $h_{23}=0$.

\section{\label{sec:models}Realization of the Textures}

In this Section we indicate models which might give rise to the 
textures found to be able to reproduce the data.
They originate either from the Froggatt--Nielsen mechanism \cite{FN} --- 
explaining the hierarchy among the non--zero matrix elements  --- 
or from more complicated flavor symmetries. As an example, we choose
the global non--Abelian discrete symmetry $D_4 \times Z_2$ which is broken at 
the electroweak scale by enlarging the Higgs sector through scalars
transforming non--trivially under the horizontal 
symmetry\footnote{Of course there are several ways to realize texture 
zeros in mass matrices, see, for instance \cite{klugscheisser}.}. A
global continuous symmetry broken at high energies is assumed to be
the origin of this discrete symmetry. 
Rather than working out the
details of the models, we merely reproduce 
the structure of the mass matrices, so that this Section should be 
understood as a proof of principle for the 
Dirac neutrino mass textures under consideration. One still faces the 
problem of explaining the relative smallness of neutrino masses with respect 
to the charged lepton masses. To achieve this, we have to rely on 
some tuning of the parameters of the models based on the discrete symmetry. 
In the model based on the Froggatt--Nielsen mechanism, we can explain the 
relative smallness by appropriate choice of charges and by working with 
two scalar fields, whose vacuum expectation values display a hierarchy.

\subsection{Mass Textures from the 
Non--Abelian Discrete Symmetry \Groupname{D}{4}
  $\times$ \Groupname{Z}{2}}

The group \Groupname{D}{4} has already been discussed as a possible
flavor symmetry by several authors 
\cite{Grimus:2003kq,Grimus:2004rj,Frigerio:2004jg,seidl}.  
Other dihedral and two--valued dihedral groups have also been 
considered in the literature \cite{Ma:2004br}. 
Here we discuss \Groupname{D}{4} for the first time in the 
context of Dirac neutrinos. 
Mathematical details of \Groupname{D}{4} are delegated to the Appendix. 

\subsubsection{Obtaining Class $A$ from 
\Groupname{D}{4} $\times$ \Groupname{Z}{2}}

Consider the 
Dirac neutrino mass matrix to be of the form of \Eqref{eq:A} and 
the charged lepton mass matrix to be diagonal. 
As shown in the following, this setup can be realized in the 
framework of the Standard Model extended by the non--Abelian flavor
symmetry \Groupname{D}{4} $\times$ \Groupname{Z}{2}. The left--handed
leptons, right--handed charged leptons and neutrinos are assigned to
be:
\[ \ba 
\left\{L_e, \left( \begin{array}{cc} L_\mu \\ L_\tau \end{array}
  \right) \right\} \sim
\left\{\MoreRep{1}{1}^{+},\Rep{2}^{+}\right\} 
\;,\;\;
\left\{ e_R, \left( \begin{array}{cc} \mu_R\\ \tau_R
    \end{array} \right) \right\} \sim
\left\{\MoreRep{1}{1}^{-},\Rep{2}^{-} \right\} \;,\;\;\\
\left\{ (\nu_e)_R, (\nu_\mu)_R, (\nu_\tau)_R \right\} \sim
\left\{\MoreRep{1}{1}^{-},\MoreRep{1}{2}^{+}, \MoreRep{1}{2}^{-} \right\} 
.
\ea
\]
The left--handed lepton doublets are 
$L_e = \left( \bad \nu_e \\ e \ea \right)_L$ 
and analogously for $L_\mu$ and $L_\tau$.  
The Higgs sector of the model contains 5 SM--like $SU(2)_L$ 
doublets, two of which 
form a two--dimensional representation under $D_4$ and the three 
other ones a one--dimensional one:   
\[ 
\phi _{1} ^{-} \sim \MoreRep{1}{1}^{-} \;,\;\; 
\phi_{2} ^{+} \sim \MoreRep{1}{2} ^{+} \;,\;\; \phi_{4} ^{-} \sim
\MoreRep{1}{4} ^{-}\;,\;\;\left( \begin{array}{c}
    \psi _{1}^{+} \\ \psi_{2}^{+} \end{array} \right) \sim \Rep{2}^{+}
~,
\]
where $\MoreRep{i}{j}$ denotes the $j^{\mathrm{th}}$ representation of
dimension $i$ of \Groupname{D}{4} and the superscript $^{\pm}$ the
transformation property under \Groupname{Z}{2}. 
Therefore, a field $\phi^+$ transforms under $Z_2$ as 
$\phi^+$, whereas $\phi^-$ transforms as $-\phi^-$. 

\noindent Using real representation matrices for the two--dimensional 
representation, this leads to the following tree level mass matrices 
\[
\ba 
m_{\nu} = \left( \begin{array}{ccc} 
    \Yuk{1} \langle \phi_{1}^{-} \rangle 
& \Yuk{2} \langle \phi _{2} ^{+} \rangle & 0\\
    0&\Yuk{3} \langle \psi _{2} ^{+} \rangle  & 0\\
    0&\Yuk{3} \langle  \psi _{1} ^{+} \rangle & 0
\end{array} \right) \;\; \mbox{and} \;\; \\
m_{\ell}= \left( \begin{array}{ccc}
    \Yuk{4} \langle \phi_{1} ^{-} \rangle & 0 & 0\\
    0&\Yuk{5} \langle \phi_{1} ^{-} \rangle 
+ \Yuk{6} \langle \phi _{4} ^{-}\rangle & 0\\
    0&0&\Yuk{5} \langle \phi_{1} ^{-} \rangle - 
\Yuk{6} \langle \phi _{4} ^{-}\rangle 
\end{array} \right)~, 
\ea
\]
which corresponds to the Dirac mass matrix of class $A$ with a diagonal 
charged lepton mass matrix. The Yukawa couplings are denoted with 
$\kappa_i$ and the Higgs fields $\phi_{i}^{\pm}$
and $\psi_{i}^{\pm}$ have acquired their vacuum expectation values 
(vevs). Obtaining a hierarchy 
for the entries of the mass matrices requires for this and the 
next two examples, in general a detailed analysis of the Higgs
potential, which is beyond the scope of the present work. Still one
can give some order--of--magnitude estimates for the vevs and Yukawa
couplings. This illustrates the tuning required to explain the relative 
smallness of neutrino and charged lepton masses. Invoking the framework 
in a scenario involving higher dimensions might explain the overall 
suppression of neutrino masses along the lines of \cite{large, otherED}. 
This is well beyond the scope of our work, which is supposed to deal 
mainly with the phenomenological consequences of texture zeros. 
Let us give an example of the required values of couplings and vevs: 
with $v$ being the electroweak scale one can choose 
$\langle \phi_{2} ^{+} \rangle = \mathcal{O}(10^{-14}) \,\,
v \,$,$ \, \langle \psi_{1} ^{+} \rangle = \mathcal{O}(10^{-12}) \,\,
v \, $,$ \, \langle \psi_{2} ^{+} \rangle = \mathcal{O}(10^{-12})
\,\, v \,$,
$ \, \langle \phi_{1} ^{-} \rangle = \mathcal{O}(10^{-2}) \,\, v \,
$, $\, \langle
\phi_{4} ^{-} \rangle = \mathcal{O}(10^{-2}) \,\, v \,$ for the vevs and
$\kappa_{1} = \mathcal{O}(10^{-10})$, $\kappa_{4} = 
\mathcal{O}(10^{-4})$, $\kappa_{2} \, $,$\, \kappa_{3} \, 
$,$\, \kappa_{6} = \mathcal{O}(1)$ and 
$\kappa_{5} =  \mathcal{O}(1)$. Note that with these choices 
there has to exist one further Higgs doublet whose vev is of the order of $v$. 
For example, one might introduce a Higgs field transforming as 
$\MoreRep{1}{1} ^{+}$, which cannot couple to the leptons 
because of their assignments under $D_4 \times Z_2$, but 
can give adequate masses to the quarks.

\subsubsection{Obtaining Class $B$ from 
\Groupname{D}{4} $\times$ \Groupname{Z}{2}}
To obtain class $B$ from the group 
\Groupname{D}{4} $\times$ \Groupname{Z}{2} 
we choose the following assignments for the leptons: 
\[ \ba
\left\{ \left( \begin{array}{c} L_e\\ L_\mu \end{array} \right) ,L_\tau \right\} \sim
\left\{\Rep{2}^{+},\MoreRep{1}{1}^{+} \right\} \;,\;\;
\left\{ \left(\begin{array}{c} e_R \\ \mu_R \end{array}
  \right),\tau_R \right\} \sim \left\{
  \Rep{2}^{-},\MoreRep{1}{1}^{-} \right\} \;,\;\; \\
\left\{ \left( \begin{array}{c}(\nu_e)_R\\ (\nu_\mu)_R \end{array}
    \right), (\nu_\tau)_R \right\} \sim \left\{\Rep{2}^{+},
\MoreRep{1}{2}^{-} \right\}~, 
\ea \]
while for the Higgs fields we now require 6 doublets:
\[
\phi_{1}^{-} \sim \MoreRep{1}{1}^{-} \;,\;\;
\phi_{2} ^{+} \sim \MoreRep{1}{2} ^{+} \;,\;\; \phi_{3} ^{+} \sim
\MoreRep{1}{3} ^{+} \;,\;\;
\phi_{4} ^{-} \sim \MoreRep{1}{4} ^{-}
\;,\;\; \; \left( \begin{array}{c}
    \psi_{1}^{+} \\ \psi_{2} ^{+} \end{array} \right) \sim \Rep{2} ^{+}~.
\]
These choices lead to the following mass matrices:
\[ 
\ba  
m_{\nu} = \left( \begin{array}{ccc} 
    0& \Yuk{2} \langle \phi _{2} ^{+} \rangle 
- \Yuk{3} \langle \phi_{3} ^{+}\rangle & 0\\
    \Yuk{2} \langle \phi_{2} ^{+} \rangle 
+ \Yuk{3} \langle \phi_{3} ^{+}\rangle &0&0\\
    \Yuk{1} \langle \psi_{1} ^{+}\rangle &
\Yuk{1} \langle \psi _{2} ^{+}\rangle & 0
\end{array} \right) \;\; \mbox{and} \;\; \\
m_{\ell}= \left( \begin{array}{ccc}
    \Yuk{4} \langle \phi_{1}^{-} \rangle 
+ \Yuk{5} \langle \phi_{4} ^{-}\rangle &0&0\\
    0&\Yuk{4} \langle \phi_{1} ^{-} \rangle  
-\Yuk{5} \langle \phi _{4} ^{-}\rangle &0\\
    0&0&\Yuk{6} \langle \phi_{1} ^{-} \rangle 
\end{array} \right)~.
\ea
\]
The hierarchy among the elements in the mass matrices can be
reached in a similar way as explained in the previous example. 
Note that here we have different vevs appearing in the matrices governing 
$m_\nu$ and $m_\ell$, which slightly simplifies the issue.  

\subsubsection{Obtaining Class $D_{1}$ from 
\Groupname{D}{4} $\times$ \Groupname{Z}{2}}

A texture belonging to class $D_{1}$ can be reproduced by the 
following assignments for the leptons which is very similar to the
assignment used to create textures belonging to the class $B$: 
\[ \ba 
\left\{ \left( \begin{array}{c} L_e\\ L_\mu \end{array} \right), 
L_\tau \right\} \sim
\left\{ \Rep{2}^{+},\MoreRep{1}{1}^{+} \right\} 
\;,\;\;
\left\{ \left( \begin{array}{c} e_R \\  \mu_R \end{array}
  \right), \tau_R \right\} \sim \left\{
  \Rep{2}^{-},\MoreRep{1}{1}^{-} \right\}\;,\;\; \\
\left\{ \left( \begin{array}{c} (\nu_e)_R \\ (\nu_\mu)_R
    \end{array} \right), (\nu_\tau)_R \right\} \sim
\left\{ \Rep{2}^{+},\MoreRep{1}{1}^{-} \right\}
\ea \]
For the Higgs particles we choose: 
\[
\phi _{1}^{+} \sim \MoreRep{1}{1} ^{+}  \;,\;\; 
\phi _{1}^{-} \sim \MoreRep{1}{1} ^{-}  \;,\;\;
\phi_{4} ^{+} \sim
\MoreRep{1}{4} ^{+} \;,\;\;
\phi_{4} ^{-} \sim \MoreRep{1}{4} ^{-}
\;,\;\;
\left( \begin{array}{c} \psi_{1} ^{+} \\
    \psi_{2} ^{+} \end{array} \right) \sim \Rep{2}^{+}\;\;\;~,
\]
yielding the mass matrices
\[ 
\ba 
m_{\nu} = \left( \begin{array}{ccc} 
    \Yuk{1} \langle \phi_{1} ^{+} \rangle 
+ \Yuk{2} \langle \phi_{4} ^{+}\rangle & 0 & 0\\
    0& \Yuk{1} \langle \phi_{1} ^{+} \rangle 
- \Yuk{2} \langle \phi_{4} ^{+}\rangle &0\\
    \Yuk{3} \langle \psi_{1} ^{+}\rangle 
& \Yuk{3} \langle \psi_{2} ^{+}\rangle & \Yuk{4} \langle \phi_{1} ^{-} \rangle 
\end{array} \right) \;\; \mbox{and} \;\; \\
m_{\ell}= \left( \begin{array}{ccc}
    \Yuk{5} \langle \phi_{1} ^{-} \rangle 
+ \Yuk{6} \langle \phi_{4} ^{-}\rangle &0&0\\
    0& \Yuk{5} \langle \phi_{1} ^{-} \rangle 
- \Yuk{6} \langle \phi_{4} ^{-} \rangle &0\\
    0&0& \Yuk{7} \langle \phi_{1} ^{-} \rangle 
\end{array} \right).
\ea
\]
Again we can reproduce the hierarchy among the elements and the 
relative smallness of the neutrino masses with an interplay of vevs and 
Yukawa couplings. 

\subsection{Mass Textures from the Froggatt--Nielsen mechanism}

In this Subsection we show how one can generate the 
textures under consideration in a
supersymmetric framework by assuming 
the existence of two global horizontal flavor symmetries denoted 
$U(1)_A$ and $U(1)_B$.  
As an example, we generate class $A$ of Section \ref{sec:5}. 
In addition to the two global $U(1)$ symmetries, 
we require the presence of the cyclic symmetry \Groupname{Z}{3} 
acting only on the charged leptons to preserve their mass matrix to be
diagonal. Under this \Groupname{Z}{3} symmetry, the right-- and 
left--handed charged leptons transform according to 
$\sim \left\{ 1, \omega, \omega^2\right\}$, where $\omega = e^{2 \pi i /3}$. 
The $U(1)_{A,B}$ are both broken at 
high--energy scales $M_{A}$ and $M_{B}$, respectively. 
There are two scalar fields $\sigma$ and $\rho$, which 
are gauge group singlets. The field $\sigma$ ($\rho$) carries 
charge $-1$ under $U(1)_A$ $(U(1)_B)$ and, by acquiring a  
vev, whose magnitude is of the order $M_A$ ($M_B$), breaks 
the horizontal symmetry spontaneously. Small parameters  
associated to this framework are given by the ratio of those vevs with some 
high scale $M$, which corresponds to the mass of the additional heavy
vector-like fermions, which are integrated out. 
The $U(1)$--charges are chosen as follows:
\[
\left\{ L_e, L_\mu, L_\tau \right\} \sim \left\{
  (3_{A},0_{B}),(5_{A},0_{B}),(5_{A},0_{B}) \right\} ~, 
\]
\[
\left\{ e_R, \mu_R, \tau_R \right\} 
\sim \left\{ (8_{A},0_{B}),(7_{A},0_{B}),(5_{A},0_{B}) \right\} ~,
\]
\[ 
\left\{ (\nu_e)_R, (\nu_\mu)_R, (\nu_\tau)_R \right\} 
\sim \left\{ (3_{A},6_{B}),(5_{A},6_{B}),(0_{A},6_{B}) \right\}~. 
\]
The MSSM Higgs doublets $H_{u}$ and $H_{d}$ are assumed to have zero charges under both $U(1)$ symmetries. 
For the fields $\rho$ and $\sigma$ we choose the following possibilities 
for their charge assignment and their vevs: 
\[
\sigma \sim (-1_{A},0_{B}) \;\; \mbox{with } \;\; 
\langle \sigma \rangle/M
\equiv \lambda \sim 0.22~, 
\]
\[
\rho \sim (0_{A},-1_{B}) \;\; \mbox{with } \;\; 
\langle \rho \rangle/M
\equiv \epsilon \sim 10^{-2} ~.
\]
The relative smallness of $\langle \rho \rangle/M$ is 
required to explain the hierarchy between the charged lepton and 
neutrino masses and between the leptons and quarks, whereas 
$\langle \sigma \rangle/M$ is 
responsible for the mass hierarchy between the generations of the 
charged leptons and neutrinos, respectively. 
We end up with:
\[
m_{\nu} = \Order{\epsilon^{6}} \left( \begin{array}{ccc} 
    \Order{1} & \Order{\lambda^{2}} & 0\\
    0&\Order{1}&0\\
    0&\Order{1}&0
\end{array} \right) \langle H_{u}\rangle \;\; \mbox{and} \;\; 
m_{\ell}= \Order{1} \left( \begin{array}{ccc}
    \Order{\lambda^{5}}&0&0\\
    0&\Order{\lambda^{2}}&0\\
    0&0&\Order{1}
\end{array} \right) \langle H_{d} \rangle~.
\]
Note that the zeros in $m_{\nu}$ are preserved from being filled,
since we are working in a SUSY framework, in which the conjugated field of
$\sigma$, which would carry the charges $(1_{A},0_{B})$, is
absent (``SUSY zeros''). Finally we would like to note that 
the relative smallness of the neutrino masses has its origin in 
an appropriate choice of the charges and the small 
value of $\langle \rho \rangle/M$ with respect to 
$\langle \sigma \rangle/M$. 

\section{\label{sec:concl}Conclusions, Summary and final Remarks}
Apart from theoretical prejudices, 
we have no incontrovertible evidence for the Majorana nature of neutrinos. 
If they are Dirac particles, just like all other known fermions, 
one might be interested 
in the form of the Dirac mass matrix $m_\nu$ implied by current data. 
In particular, the minimal allowed structure of $m_\nu$ might be of 
special interest. Here, the situation is analyzed in terms of putting as 
many zeros in $m_\nu$ as possible. 
As a proof of principle, we have also outlined possibilities to obtain the 
successful Dirac neutrino mass matrix textures from models based on either 
discrete non--Abelian symmetries or on a Froggatt--Nielsen scenario. 
We found that up to 5 zero entries in $m_\nu$ are allowed 
by current data. All of those candidates imply $CP$ conservation, 
one massless neutrino and their phenomenology is summarized in Table 
\ref{tab:tab5yes}. 
Most successful matrices with 4 zero entries also 
predict $CP$ conservation, their physical implications are summarized in 
Tables \ref{tab:tab4yes1} and \ref{tab:tab4yes2}. 
Apart from a few exceptions, which then reproduce the 
phenomenology of already discussed matrices with 5 or 4 zeros, 
all matrices having 3 or less zeros imply no correlation between the neutrino 
mixing observables. 
Typically the inverted hierarchy requires that there is a spread between 
the non--zero entries in $m_\nu$ of two orders of magnitude, 
whereas the normal 
hierarchy can be successfully obtained with a rather mild hierarchy. 
We would like to remark that some matrices 
predict $\theta_{13}=0$, but none of them implies in general 
$\theta_{23}= \pi/4$. This would require typically that 
two of the non--zero entries in $m_\nu$ are identical. 
It is finally interesting to note that there are basically only three 
significantly different predictions coming from Dirac mass 
matrices with zero entries.  
Those are the inverted hierarchy with zero $U_{e3}$ and the two 
possibilities given in Eqs.\ (\ref{eq:12zero},\ref{eq:13zero}) and 
(\ref{eq:EcorNH},\ref{eq:EcorIH}). Note that once there is an interesting 
correlation between observables, $CP$ is automatically conserved. 
The characteristic value of 
$U_{e3}$ which can play the role of a discriminator between those 
possibilities, is given by 
$|U_{e3}| \simeq  \frac{1}{2} R \sin 2 \theta_{12}$.

\vspace{0.5cm}
\begin{center}
{\bf Acknowledgments}
\end{center}
This work was supported by the ``Deutsche Forschungsgemeinschaft'' in the 
``Sonderforschungsbereich 375 f\"ur Astroteilchenphysik'' (C.H.\ and W.R.) 
and under project number RO-2516/3-1 (W.R.).

\begin{table}\hspace{-.7cm}
\begin{tabular}{|c|c|c|p{1.7in}|c|c|}
\hline 
Class & $N$ & Condition & Phenomenology & 
\multicolumn{2}{c|}{Order of magnitude}\\
& & & & NH & IH\\ 
\hline \hline
\rule[0.4in]{0cm}{0cm}$
A = 
\left( 
\begin{array}{ccc} 
a & b & 0  \\
0 & d & 0 \\
0 & e & 0 
\end{array}
\right)$&6&$\left| \begin{array}{cc}  h_{22}&h_{23}\\h_{32}& h_{33}
  \end{array} \right|=0$& \vspace{-.6cm}
only IH \newline 
$U_{e3}=0$,  $J_{CP}=0$
&---&$\left(\begin{matrix}  
 1&\lambda^2&0\\
  0&1&0\\
  0&1&0 
\end{matrix} \right)$\\[0.25in]
\hline \hline
\rule[0.4in]{0cm}{0cm}$B=\left(\begin{array}{ccc} 
0 & a & 0 \\
b & 0 & 0 \\
d & e & 0 
\end{array} \right)$ & 6 
& $h_{12}=0$ 
& \vspace{-.9cm}
$J_{CP}=0$ \newline 
$U_{e3} \simeq \frac{R}{2}\sin 2 \theta_{12}\,  \cot \theta_{23}$ \newline 
Eq.\ (\ref{eq:12zero}), Fig.\ \ref{fig:fig1a} 
&$\left(\begin{matrix}  
 0&\lambda&0\\
 1&0&0\\
  1&\lambda&0 
\end{matrix} \right)$ &$\left(\begin{matrix}  
 0&1&0\\
 1&0&0\\
  1&\lambda^2&0 
\end{matrix} \right)$\\[0.25in] 
\hline
\rule[0.4in]{0cm}{0cm}$\tilde{B}=\left(\begin{array}{ccc} 
0 & a & 0 \\
b & d & 0 \\
e & 0 & 0 
\end{array} \right)$&6&$h_{13}=0$& \vspace{-.9cm}
$J_{CP}=0$ \newline 
$U_{e3} \simeq \frac{R}{2}\sin 2 \theta_{12} \, \tan \theta_{23}$\newline 
Eq.\ (\ref{eq:13zero}), Fig.\ \ref{fig:fig1}
&$\left(\begin{matrix}  
 0&\lambda&0\\
 1&\lambda&0\\
  1&0&0 
\end{matrix} \right)$ &$\left(\begin{matrix}  
 0&1&0\\
 1&\lambda^2&0\\
  1&0&0 
\end{matrix} \right)$\\[0.25in]
\hline
\end{tabular}
\caption{\label{tab:tab5yes}
Successful matrices with five zero entries. Shown are a typical 
representants of the class, the number $N$ of distinct matrices in that 
class, the characteristic condition giving its phenomenology and 
an order of magnitude estimate of its entries. We also indicate from which 
equation the phenomenology results. 
NH denotes normal hierarchy and IH inverted hierarchy. Note that 
the smallest mass state is always zero because the determinant of 
$m_\nu$ vanishes.} 
\end{table}

\begin{table}
\begin{center}
\rule[0in]{-1.3cm}{0cm}\begin{tabular}{|c|c|c|p{1.2in}|c|c|}
\hline 
Class&$N$&Condition&Phenomenology&\multicolumn{2}{c|}{Order of magnitude}\\
&&&&NH&IH\\
\hline \hline
\rule[0.4in]{0cm}{0cm}$C=\left( \begin{array}{ccc}
  a&b&d\\
  e&0&0\\
  f&0&0
\end{array} \right)$&3&$\left| \begin{array}{cc}
  h_{22}&h_{23}\\h_{32}& h_{33} \end{array} \right|=0$&identical to $A$&---& $\left(\begin{matrix} 
  \lambda^2&1&\lambda^2\\
  1&0&0\\
  1&0&0
\end{matrix} \right)$\\[0.25in]
\hline \hline
\rule[0.4in]{0cm}{0cm}$F_{1}=\left(\begin{array}{ccc}
    a&0&0\\
    b&d&0\\
    e&f&0
    \end{array} \right)$&6&---& 
\vspace{-.6cm}$J_{CP} \neq 0$ \newline $U_{e3} \neq 0$
&$\left(\begin{matrix}
    \lambda&0&0\\
    \lambda&1&0\\
    \lambda&1&0
    \end{matrix} \right)$&$\left(\begin{matrix}
    1&0&0\\
    \lambda^2&1&0\\
    \lambda^2&1&0
    \end{matrix} \right)$\\[0.25in]
\hline
\rule[0.4in]{0cm}{0cm}$F_{2}=\left(\begin{array}{ccc}
    a&b&0\\
    d&0&0\\
    e&f&0
    \end{array} \right)$&6&---& 
same as above 
&$\left(\begin{matrix}
    \lambda^2&\lambda&0\\
    1&0&0\\
    1&\lambda&0
    \end{matrix} \right)$&$\left(\begin{matrix}
    \lambda^2&1&0\\
    1&0&0\\
    1&\lambda^2&0
    \end{matrix} \right)$\\[0.25in]
\hline
\rule[0.4in]{0cm}{0cm}$F_{3}=\left(\begin{array}{ccc}
    a&b&0\\
    d&e&0\\
    f&0&0
    \end{array} \right)$&6&---& 
same as above 
&$\left(\begin{matrix}
    \lambda^2&\lambda&0\\
    1&\lambda&0\\
    1&0&0
    \end{matrix} \right)$&$\left(\begin{matrix}
    \lambda^2&1&0\\
    1&\lambda^2&0\\
    1&0&0
    \end{matrix} \right)$\\[0.25in]
\hline 
\end{tabular}
\caption{\label{tab:tab4yes1}
Same as Table \ref{tab:tab5yes} for matrices with four zero entries 
and one vanishing mass eigenvalue.} 
\end{center}
\end{table}

\begin{table}
\begin{center}
\rule[0in]{-1.3cm}{0cm}
\begin{tabular}{|c|c|c|p{1.7in}|c|c|}
\hline 
Class&$N$&Condition&Phenomenology&\multicolumn{2}{c|}{Order of magnitude}\\
&&&&NH&IH\\
\hline \hline
\rule[0.4in]{0cm}{0cm}$D_{1}= \left( \begin{array}{ccc} 
    a&0&0\\
    0&b&0\\
    d&e&f\\
\end{array} \right)$&6&$h_{12}=0$& \vspace{-.8cm}
similar to  $B$ \newline 
$U_{e3} \simeq \frac{R}{2}\sin 2 \theta_{12}\,  \cot \theta_{23}$ \newline 
Eq.\ (\ref{eq:12zero}), Fig.\ \ref{fig:fig1a} &$\left( \begin{matrix} 
    \lambda&0&0\\
    0&1&0\\
    \lambda&1&\lambda\\
\end{matrix} \right)$&$\left( \begin{matrix} 
    1&0&0\\
    0&1&0\\
    \lambda^2&1&\lambda^2\\
\end{matrix} \right)$\\[0.25in]
\hline
\rule[0.4in]{0cm}{0cm}$D_{2}= \left( \begin{array}{ccc} 
    a&0&0\\
    0&b&d\\
    e&0&f\\
\end{array} \right)$&6&$h_{12}=0$&
same as above
&$\left( \begin{matrix} 
    \lambda&0&0\\
    0&\lambda&1\\
    \lambda&0&1\\
\end{matrix} \right)$&$\left( \begin{matrix} 
    1&0&0\\
    0&\lambda^2&1\\
    \lambda^2&0&1\\
\end{matrix} \right)$\\[0.25in]
\hline
\rule[0.4in]{0cm}{0cm}$D_{3}= \left( \begin{array}{ccc} 
    a&b&0\\
    0&0&d\\
    e&0&f\\
\end{array} \right)$&6&$h_{12}=0$&
same as above 
&$ \left( \begin{matrix} 
    \lambda& \lambda & 0\\
    0&0&1\\
    \lambda& 0 & 1\\
\end{matrix} \right)$&$ \left( \begin{matrix} 
    \lambda & 1 & 0\\
    0 & 0 & 1\\
    \lambda & 0 & 1 \\
\end{matrix} \right)$\\[0.25in]
\hline \hline
\rule[0.4in]{0cm}{0cm}$\tilde{D}_{1}= \left( \begin{array}{ccc} 
    a&0&0\\
    b&d&e\\
    0&f&0\\
\end{array} \right)$&6&$h_{13}=0$&\vspace{-.8cm}
similar to  $\tilde{B}$\newline 
$U_{e3} \simeq \frac{R}{2}\sin 2 \theta_{12}\,  \tan \theta_{23}$ \newline 
Eq.\ (\ref{eq:13zero}), Fig.\ \ref{fig:fig1}&$ \left( \begin{matrix} 
    \lambda&0&0\\
    \lambda&1&\lambda\\
    0&1&0\\
\end{matrix} \right)$&$ \left( \begin{matrix} 
    1&0&0\\
    \lambda^2&1&\lambda^2\\
    0&1&0\\
\end{matrix} \right)$\\[0.25in]
\hline
\rule[0.4in]{0cm}{0cm}$\tilde{D}_{2}= \left( \begin{array}{ccc} 
    a&0&0\\
    b&d&0\\
    0&e&f\\
\end{array} \right)$&6&$h_{13}=0$&
same as above
&$ \left( \begin{matrix} 
    \lambda&0&0\\
    \lambda&1&0\\
    0&1&\lambda\\
\end{matrix} \right)$&$ \left( \begin{matrix} 
    1&0&0\\
    \lambda^2&1&0\\
    0&1&\lambda^2\\
\end{matrix} \right)$\\[0.25in]
\hline
\rule[0.4in]{0cm}{0cm}$\tilde{D}_{3}= \left( \begin{array}{ccc} 
    a&b&0\\
    d&0&e\\
    0&0&f\\
\end{array} \right)$&6&$h_{13}=0$&
same as above
&$\left( \begin{matrix} 
    \lambda&\lambda&0\\
    \lambda&0&1\\
    0&0&1\\
\end{matrix} \right)$&$\left( \begin{matrix} 
    1&\lambda&0\\
    \lambda^2&0&1\\
    0&0&1\\
\end{matrix} \right)$\\[0.25in]
\hline \hline
\rule[0.4in]{0cm}{0cm}$E=\left( \begin{array}{ccc}
    a&b&0\\
    d&0&0\\
    e&0&f
\end{array} \right)$&6&  $\frac{\D h_{12}}{\D h_{13}}=
\frac{\D h_{22}}{\D h_{23}}$
& \vspace{-.8cm} 
NH: $m_{1} \gs 0.01 \eV$ and\newline 
$U_{e3} \ge \frac{R}{2}\sin 2 \theta_{12}\,  \tan \theta_{23}$ 
\newline Eqs.\ (\ref{eq:EcorNH},\ref{eq:a1}), Fig.\ \ref{fig:fig4} \newline 
----------------------------
\newline IH: masses free and \newline
$U_{e3} \le \frac{R}{2}\sin 2 \theta_{12}\,  \tan \theta_{23}$
\newline Eqs.\ (\ref{eq:EcorIH},\ref{eq:a2}), Fig.\ \ref{fig:fig4} 
& 
$\left(\begin{matrix}
    \lambda^2&\lambda&0\\
    1&0&0\\
    1&0&\lambda^2
\end{matrix} \right)$&$\left( \begin{matrix}
    \lambda^2&1&0\\
    1&0&0\\
    1&0&\lambda^2
\end{matrix} \right)$\\[0.25in]
\hline \hline
\rule[0.4in]{0cm}{0cm}$\tilde{E}=\left( \begin{array}{ccc}
    a&0&b\\
    d&e&0\\
    f&0&0
\end{array} \right)$&6&$\frac{\D h_{12}}{\D h_{13}}=\frac{\D h_{32}}{\D h_{33}}$&\vspace{-.8cm}  as class $E$ with \newline
$\theta_{23} \rightarrow \theta_{23} + \pi/4$&$\left( \begin{matrix}
    \lambda^2&0&\lambda\\
    1&\lambda^2&0\\
    1&0&0
\end{matrix} \right)$&$\left( \begin{matrix}
    \lambda^2&0&1\\
    1&\lambda^2&0\\
    1&0&0
\end{matrix} \right)$\\[0.25in]
\hline 
\end{tabular}
\caption{\label{tab:tab4yes2}Same as Table \ref{tab:tab4yes1} 
for matrices with four zero entries 
and no vanishing mass eigenvalue. All classes predict $J_{CP}=0$.}
\end{center}
\end{table}

\begin{table}
\begin{center}
\begin{tabular}{|c|c|c|}
\hline 
Symmetric Class from Eq.\ (\ref{eq:G}) & Results of 
\cite{2zeros0} & Order of magnitude  \\
\hline \hline
$G_{1} = \left(\begin{array}{ccc}
    0&0&a\\
    0&b&d\\
    a&d&e
\end{array}\right)$& $ \ba {\rm Case~} A_{1} \\ 
\mbox{only NH} \\
|U_{e3}| \simeq \frac{\sqrt{R}}{2} \cot \theta_{23} 
\frac{\sin 2 \theta_{12}}{ \sqrt{\cos 2 \theta_{12}}}  \ea $ & 
$\left( \begin{matrix}
    0 & 0 & \lambda\\
    0 & 1 & 1\\
    \lambda & 1 & 1
\end{matrix} \right)$
\\
\hline
$G_{2} = \left(\begin{array}{ccc}
    0&a&0\\
    a&b&d\\
    0&d&e
\end{array}\right)$&
$ \ba {\rm Case~} A_{2} \\ 
\mbox{only NH} \\
|U_{e3}| \simeq \frac{\sqrt{R}}{2} \tan \theta_{23} 
\frac{\sin 2 \theta_{12}}{ \sqrt{\cos 2 \theta_{12}}}  \ea $ & 
$\left( \begin{matrix}
    0 & \lambda & 0 \\
    \lambda  & 1 & 1\\
    0 & 1 & 1
\end{matrix} \right)$
 \\
\hline \hline
$G_{3}=\left(\begin{array}{ccc}
    a&0&b\\
    0&0&d\\
    b&d&e
\end{array}\right)$& $ \ba {\rm Case~} B_{3} \\ \mbox{QD; 
both orderings} \\
|U_{e3}| \simeq R \, |\frac{\cot 2\theta_{23} }{\cos \delta }|
\frac{\tan\theta_{12}}{1 + \tan^2 \theta_{12}} 
\ea $ & $\left( \begin{matrix}
    1 & 0 & \lambda^2 \\
    0 & 0 & 1 \\
    \lambda^2 & 1 & \lambda 
\end{matrix} \right)$
\\
\hline
$G_{7}=\left(\begin{array}{ccc}
    a&b&0\\
    b&d&e\\
    0&e&0
\end{array}\right)$& $ \ba {\rm Case~} B_{4} \\ \mbox{QD; 
both orderings} \\
|U_{e3}| \simeq R \, |\frac{\tan 2\theta_{23} }{\cos \delta }|
\frac{\tan\theta_{12}}{1 + \tan^2 \theta_{12}} 
\ea $ & $\left( \begin{matrix}
    1 & \lambda^2 & 0 \\
    \lambda^2 & \lambda & 1 \\
    0 & 1 & 0  
\end{matrix} \right)$ \\
\cline{2-3} 
\hline
$G_{10}=\left(\begin{array}{ccc}
    a&0&b\\
    0&d&e\\
    b&e&0
\end{array}\right)$& $ \ba {\rm Case~} B_{2} \\ 
\mbox{same as } 
B_4~(\mbox{or } G_7) \ea $ &  $\left( \begin{matrix}
    1 & 0 & \lambda^2 \\
    0 & \lambda & 1 \\
    \lambda^2 & 1 & 0  
\end{matrix} \right)$\\
\cline{2-3}
and~~~ $\left(\begin{array}{ccc}
    a&b&0\\
    b&0&d\\
    0&d&e
\end{array}\right)$& $ \ba {\rm Case~} B_{1} \\ \mbox{same as } 
B_3~(\mbox{or } G_3) \ea $ & $\left( \begin{matrix}
    1 & \lambda^2 & 0 \\
    \lambda^2 & 0 & 1 \\
    0 & 1 & \lambda 
\end{matrix} \right)$
\\
\hline \hline 
$\left(\begin{array}{ccc}
    a&b&d\\
    b&0&e\\
    d&e&0
\end{array}\right)$ & $\ba {\rm Case~C} \\ 
\mbox{QD; 
both orderings} \\ |U_{e3}| \simeq 
\frac{\cot 2 \theta_{12} \, \cot 2 \theta_{23} }{\cos \delta }
\ea $  & $\left( \begin{matrix}
    1 & \lambda^2 & \lambda \\
    \lambda^2 & 0 & 1 \\
    \lambda & 1 & 0  
\end{matrix} \right)$
\\ \hline  
\end{tabular}
\caption{\label{tab:FGM}Symmetric Dirac mass matrices from 
Eq.\ (\ref{eq:G}) together with the corresponding two zero Majorana mass 
matrices from Refs.\ \cite{2zeros0}. Class $C$ from 
\cite{2zeros0} would correspond to a symmetric Dirac 
mass matrix with two zero entries.}
\end{center}
\end{table}

\begin{table}
\begin{center}
\begin{tabular}{|c|c|c|p{1.7in}|}
\hline 
Class&$N$&Condition&Phenomenology\\
\hline \hline
\multicolumn{4}{|c|}{matrices with one vanishing eigenvalue}\\
\hline
\rule[0.4in]{0cm}{0cm}$
\left( 
\bad 
0 & b & 0  \\
0 & d & 0 \\
a & e & 0 
\ea
\right)$&6&$\left| \begin{array}{cc} h_{11}&h_{12}\\h_{21}& h_{22}
  \end{array} \right|=0$& \vspace{-0.6cm} NH: $U_{\tau 1}=0$ \newline IH: $U_{\tau 3}=0$\\[0.1in]
\hline \hline
\rule[0.4in]{0cm}{0cm}$\left(\begin{array}{ccc}
a & b & 0 \\
d & 0 & 0 \\
0 & e & 0 
\end{array} \right)$&6&$h_{23}=0$& 
$|U_{e3}|$ too large 
\\[0.15in]
\hline \hline
\rule[0.4in]{0cm}{0cm}$\left( \begin{array}{ccc}
  a&0&0\\
  b&d&e\\
  f&0&0
\end{array} \right)$&3&$\left| \begin{array}{cc} h_{11}&h_{13}\\h_{31}& h_{33} \end{array} \right|=0$& 
 \vspace{-0.6cm} NH: $U_{\mu 1}=0$ \newline IH: $U_{\mu 3}=0$
\\[0.25in]
\hline \hline
\multicolumn{4}{|c|}{matrices with all $m_{i} \neq 0$}\\
\hline
\rule[0.4in]{0cm}{0cm}$ \left( \begin{array}{ccc} 
    a&b&d\\
    0&0&e\\
    0&f&0\\
\end{array} \right)$&6&$h_{23}=0$& 
\vspace{-0.3cm}similar to second matrix \newline in this Table \\[0.25in]
\hline
\rule[0.4in]{0cm}{0cm}$\left( \begin{array}{ccc}
    a&0&0\\
    b&0&d\\
    e&f&0
\end{array}
\right)$&6&$\frac{\D h_{11}}{\D h_{12}}=\frac{\D h_{31}}{\D h_{32}}$& 
$|U_{e3}|$ too large 
\\[0.25in]
\hline 
\end{tabular}
\caption{\label{tab:dontwork}Typical matrices with four and five zeros 
unable to accommodate the data. For discussion see 
Section \ref{sec:dontwork}.}
\end{center}
\end{table}

\newpage
\begin{appendix}
\section{Group theory of \Groupname{D}{4}}

The group theory is given by the following character table,
generator relations and matrices \cite{books}: 
\begin{center}
\begin{tabular}{l|ccccc}
&\multicolumn{5}{c}{classes}                                                  \\ \cline{2-6}
&$\Cl{1}$&$\Cl{2}$&$\Cl{3}$&$\Cl{4}$&$\Cl{5}$\\
\cline{1-6}
\rule[0.15in]{0cm}{0cm} $\rm G$         &$\rm \mathbb{1}$&$\rm A ^{2}$  &$\rm B$        &$\rm A B$      &$\rm A$\\
\cline{1-6}
$\OrdCl{i}$          &1      &1      &2      &2      &2\\
\cline{1-6}
$\Ord{h}{} _{\Cl{i}}$                   &1      &2      &2      &2      &4             \\
\hline
$\MoreRep{1}{1}$                        &1      &1      &1      &1      &1                             \\
$\MoreRep{1}{2}$                        &1      &1      &-1     &1      &-1                            \\
$\MoreRep{1}{3}$                        &1      &1      &-1     &-1     &1                             \\
$\MoreRep{1}{4}$                        &1      &1      &1      &-1     &-1                            \\
$\Rep{2}$                               &2      &-2     &0      &0      &0                             \\
\end{tabular}
\end{center}
Here we denoted with $\MoreRep{1}{1,2,3,4}$ and $\Rep{2}$ the irreducible 
representations of \Groupname{D}{4},  
$G$ are the generators of the group, 
$\OrdCl{i}$ is the order of the class $\Cl{i}$ and 
$\Ord{h}{}_{\Cl{i}}$ is the order of the elements in class $\Cl{i}$, 
where $i=1,\ldots,5$.  
The generators $\rm A$ and $\rm B$ for $\Rep{2}$ are 
\[
\rm A=\left(\begin{array}{cc} 
                                              0  & -1 \\ 
                                              1  &  0
                                \end{array}\right)  
\;\;\; \mbox{and} \;\;\; \rm B=\left(\begin{array}{cc} 
                                                1 & 0 \\
                                                0 & -1   
                                \end{array}\right)~,
\]
and fulfill the generator relations 
\[
\rm A^4=\mathbb{1}\;\;\; , \;\;\; B^2=\mathbb{1} \;\;\;
\mbox{and} \;\;\; ABA=B~. 
\]
From the character table one can read off the Kronecker products:
\begin{center}
\begin{tabular}{c|cccc}
$\times$&$\MoreRep{1}{1}$&$\MoreRep{1}{2}$&$\MoreRep{1}{3}$&$\MoreRep{1}{4}$\\
\hline
$\MoreRep{1}{1}$        &$\MoreRep{1}{1}$       &$\MoreRep{1}{2}$       &$\MoreRep{1}{3}$       &$\MoreRep{1}{4}$\\
$\MoreRep{1}{2}$        &$\MoreRep{1}{2}$       &$\MoreRep{1}{1}$       &$\MoreRep{1}{4}$       &$\MoreRep{1}{3}$\\
$\MoreRep{1}{3}$        &$\MoreRep{1}{3}$       &$\MoreRep{1}{4}$       &$\MoreRep{1}{1}$       &$\MoreRep{1}{2}$\\
$\MoreRep{1}{4}$        &$\MoreRep{1}{4}$       &$\MoreRep{1}{3}$       &$\MoreRep{1}{2}$       &$\MoreRep{1}{1}$\\
\end{tabular}
\end{center}
\[
\Rep{2} \times \MoreRep{1}{i}= \Rep{2} \;\;\; \forall \; \rm i~,~
\left[ \Rep{2} \times \Rep{2}\right] = \MoreRep{1}{1} + \MoreRep{1}{2} + \MoreRep{1}{4} ~,~ \left\{ \Rep{2} \times \Rep{2}\right\} = \MoreRep{1}{3}~, 
\]
where $\left[ \Rep{n} \times \Rep{n} \right]$ denotes the symmetric
and $\left\{ \Rep{n} \times \Rep{n} \right\}$ the anti--symmetric part
of the product $\Rep{n} \times \Rep{n}$. 

\noindent The Clebsch--Gordan coefficients for the products $\Rep{2} \times
\MoreRep{1}{i}$ are then with $\left( \begin{array}{c} s_{1}\\
  s_{2}\end{array} \right) \sim \Rep{2}$ and $t \sim \MoreRep{1}{i}$:\\
\begin{center}
\begin{tabular}{ccccc}
$i=1$&$i=2$&$i=3$&$i=4$&\\[0.1in]
$\left(\begin{array}{c} s_{1} t \\ s_{2} t \end{array}
\right)$&$\left(\begin{array}{c} s_{2} t \\ s_{1} t \end{array}
\right)$&$\left(\begin{array}{c} s_{2} t \\ -s_{1} t \end{array}
\right)$&$\left(\begin{array}{c} s_{1} t \\ -s_{2} t \end{array}
\right)$& $\sim \Rep{2}$\\
\end{tabular}
\end{center}
and for the product $\Rep{2} \times \Rep{2}$ for $\left(\begin{array}{c}
  s_{1} \\ s_{2} \end{array} \right), \; \left(\begin{array}{c}
  t_{1} \\ t_{2} \end{array} \right) \sim \Rep{2}$:
\[
(s_{1} t_{1} + s_{2} t_{2})/\sqrt{2} \sim \MoreRep{1}{1} \;\;\;
\mbox{and} \;\;\;  (s_{1} t_{2} + s_{2} t_{1})/\sqrt{2} \sim \MoreRep{1}{2}~,
\]
\[
(s_{1} t_{2} -s_{2} t_{1})/\sqrt{2} \sim \MoreRep{1}{3} \;\;\;
\mbox{and} \;\;\;  (s_{1} t_{1} -s_{2} t_{2})/\sqrt{2} \sim \MoreRep{1}{4}~.
\]

\end{appendix}


\begin{thebibliography}{99} 


\bibitem{reviews}
G.~Altarelli and F.~Feruglio,
  New J.\ Phys.\  {\bf 6}, 106 (2004); 
R.~N.~Mohapatra {\it et al.},
hep-ph/0412099; 
S.~T.~Petcov, Talk given at the 21st International 
Conference on Neutrino Physics and Astrophysics (Neutrino 2004), 
June 14 - 19, 2004, Paris, France, 
hep-ph/0412410. 



\bibitem{APScosmo}
  S.~W.~Barwick {\it et al.},
  astro-ph/0412544.


\bibitem{APS0vbb}C.~Aalseth {\it et al.},
hep-ph/0412300.

\bibitem{seesaw}P. Minkowski, Phys. Lett. {\bf B67}, 421 (1977); 
T. Yanagida, in {\it Proceedings of the Workshop on
Unified Theory and the Baryon Number of the Universe}, edited by
O. Sawada and A. Sugamoto (KEK, Tsukuba, 1979), p. 95;
M. Gell-Mann, P. Ramond, and R. Slansky, in {\it Supergravity},
edited by F. van Nieuwenhuizen and D. Freedman (North Holland,
Amsterdam, 1979), p. 315;
S.L. Glashow, in {\it Quarks and Leptons}, edited by
M. L$\rm\acute{e}vy$ {\it et al.} (Plenum, New York, 1980), p. 707;
R.N. Mohapatra and G. Senjanovic, Phys. Rev. Lett. {\bf 44}, 912 (1980).






\bibitem{large}N.~Arkani-Hamed {\it et al.}, 
  Phys.\ Rev.\ D {\bf 65}, 024032 (2002); 
K.~R.~Dienes, E.~Dudas and T.~Gherghetta,
  Nucl.\ Phys.\ B {\bf 557}, 25 (1999).



\bibitem{otherED}Y.~Grossman and M.~Neubert,
  Phys.\ Lett.\ B {\bf 474}, 361 (2000); 
T.~Gherghetta,
  Phys.\ Rev.\ Lett.\  {\bf 92}, 161601 (2004); 
N.~Arkani-Hamed and M.~Schmaltz,
  Phys.\ Rev.\ D {\bf 61}, 033005 (2000); 
G.~Barenboim {\it et al.}, 
  Phys.\ Rev.\ D {\bf 64}, 073005 (2001)
P.~Q.~Hung,
  Phys.\ Rev.\ D {\bf 67}, 095011 (2003). 

\bibitem{ays}A.~Y.~Smirnov, Talk given at SEESAW25: International 
Conference on the Seesaw Mechanism and the Neutrino Mass, 
Paris, France, 10-11 Jun 2004, 
  hep-ph/0411194; 
%
A.~de Gouvea,
Mod.\ Phys.\ Lett.\ A {\bf 19} (2004) 2799. 

\bibitem{mura}H.~Murayama,
  Nucl.\ Phys.\ Proc.\ Suppl.\  {\bf 137}, 206 (2004). 

\bibitem{sugraothers}
  S.~Abel, A.~Dedes and K.~Tamvakis,
  Phys.\ Rev.\ D {\bf 71}, 033003 (2005); 
H.~Davoudiasl, R.~Kitano, G.~D.~Kribs and H.~Murayama,
  hep-ph/0502176.

\bibitem{mova}R.~N.~Mohapatra and J.~W.~F.~Valle,
  Phys.\ Rev.\ D {\bf 34}, 1642 (1986).

\bibitem{string}J.~Giedt {\it et al.}, 
  hep-th/0502032; 
for small Yukawa couplings in string theories, see also 
O.~J.~Eyton-Williams and S.~F.~King, hep-ph/0502156; 
S.~Antusch, O.~J.~Eyton-Williams and S.~F.~King,
hep-ph/0505140.

\bibitem{loops}
G.~C.~Branco and G.~Senjanovic,
  Phys.\ Rev.\ D {\bf 18}, 1621 (1978); 
D.~Chang and R.~N.~Mohapatra,
  Phys.\ Rev.\ Lett.\  {\bf 58}, 1600 (1987); 
P.~Q.~Hung,
  Phys.\ Rev.\ D {\bf 59}, 113008 (1999); 
hep-ph/0006355. 




\bibitem{other}
J.~I.~Silva-Marcos,
  Phys.\ Rev.\ D {\bf 59}, 091301 (1999); 
I.~Gogoladze and A.~Perez-Lorenzana,
  Phys.\ Rev.\ D {\bf 65}, 095011 (2002); 
Z.~Chacko {\it et al.}, 
  Phys.\ Rev.\ D {\bf 70}, 085008 (2004); 
C.~I.~Low,
Phys.\ Rev.\ D {\bf 70}, 073013 (2004); 
H.~Davoudiasl {\it et al.}, 
hep-ph/0502176.


\bibitem{lepto}M.~Fukugita and T.~Yanagida, 
Phys.\ Lett.\ B {\bf 174}, 45 (1986).

\bibitem{DiracYB}
K.~Dick {\it et al.}, 
Phys.\ Rev.\ Lett.\  {\bf 84}, 4039 (2000); 
H.~Murayama and A.~Pierce,
Phys.\ Rev.\ Lett.\  {\bf 89}, 271601 (2002). 

\bibitem{DiracYB1}
E.~K.~Akhmedov, V.~A.~Rubakov and A.~Y.~Smirnov,
Phys.\ Rev.\ Lett.\  {\bf 81}, 1359 (1998). 

\bibitem{2zeros0}
P.~H.~Frampton, S.~L.~Glashow and D.~Marfatia,
Phys.\ Lett.\ B {\bf 536}, 79 (2002); 
Z.~Z.~Xing,
Phys.\ Lett.\ B {\bf 530}, 159 (2002); 
B.~R.~Desai, D.~P.~Roy and A.~R.~Vaucher,
Mod.\ Phys.\ Lett.\ A {\bf 18}, 1355 (2003);  
W.~L.~Guo and Z.~Z.~Xing,
 Phys.\ Rev.\ D {\bf 67}, 053002 (2003). 

\bibitem{2zeroseesaw}
P.~H.~Frampton, M.~C.~Oh and T.~Yoshikawa,
  Phys.\ Rev.\ D {\bf 66}, 033007 (2002); 
A.~Kageyama {\it et al.}, 
Phys.\ Lett.\ B {\bf 538}, 96 (2002); 
S.~Kaneko, M.~Katsumata and M.~Tanimoto,
  JHEP {\bf 0307}, 025 (2003). 

\bibitem{2zerosRGE}G.~Bhattacharyya, A.~Raychaudhuri and A.~Sil,
  Phys.\ Rev.\ D {\bf 67}, 073004 (2003); 
M.~Honda, S.~Kaneko and M.~Tanimoto,
  JHEP {\bf 0309}, 028 (2003); 
C.~Hagedorn, J.~Kersten and M.~Lindner,
Phys.\ Lett.\ B {\bf 597}, 63 (2004). 

\bibitem{PMNS}B. Pontecorvo,
 Sov.\ Phys.\ JETP {\bf 6}, 429 (1957)
[Zh.\ Eksp.\ Teor.\ Fiz.\  {\bf 33}, 549 (1957)]; 
Sov.\ Phys.\ JETP {\bf 7}, 172 (1958)
[Zh.\ Eksp.\ Teor.\ Fiz.\  {\bf 34}, 247 (1957)]; 
Z. Maki, M. Nakagawa and S. Sakata, 
Prog.\ Theor.\ Phys.\  {\bf 28}, 870 (1962). 



\bibitem{valle}M.~Maltoni {\it et al.},  
New J.\ Phys.\  {\bf 6}, 122 (2004), hep-ph/0405172v4.


\bibitem{Majpha}S.~M.~Bilenky, J.~Hosek and S.~T.~Petcov,
Phys.\ Lett.\ B {\bf 94}, 495 (1980);
J.~Schechter, J.~W.~F.~Valle, 
Phys.\ Rev.\  {\bf D22}, 2227 (1980);
M.~Doi {\it et al.}, 
Phys.\ Lett.\ B {\bf 102}, 323 (1981).

\bibitem{JCP}G.~C.~Branco {\it et al.}, 
Phys.\ Rev.\ D {\bf 67}, 073025 (2003); 
see also  
G.~C.~Branco and M.~N.~Rebelo, 
New J.\ Phys.\  {\bf 7}, 86 (2005). 

\bibitem{ichPRD}W.~Rodejohann,
Phys.\ Rev.\ D {\bf 69}, 033005 (2004).


\bibitem{STP0vbb}S.~T.~Petcov,
  New J.\ Phys.\  {\bf 6}, 109 (2004).


\bibitem{DL2}W.~Rodejohann,
  Phys.\ Rev.\ D {\bf 62}, 013011 (2000); 
J.\ Phys.\ G {\bf 28}, 1477 (2002); 
K.~Zuber,
hep-ph/0008080; 
C.~S.~Lim, E.~Takasugi and M.~Yoshimura, hep-ph/0411139; 
A.~Atre, V.~Barger and T.~Han,
hep-ph/0502163.





\bibitem{MLD}M.~Lindner, M.~Ratz, M.~A.~Schmidt, to appear; 
M.~Lindner, M.~A.~Schmidt, private communication.
 



\bibitem{FN}C.~D.~Froggatt and H.~B.~Nielsen,
  Nucl.\ Phys.\ B {\bf 147}, 277 (1979).





\bibitem{klugscheisser}
W.~Grimus {\it et al.},  
  Eur.\ Phys.\ J.\ C {\bf 36}, 227 (2004). 






\bibitem{Grimus:2003kq}
  W.~Grimus and L.~Lavoura,
  Phys.\ Lett.\ B {\bf 572}, 189 (2003). 

\bibitem{Grimus:2004rj}
  W.~Grimus, A.~S.~Joshipura, S.~Kaneko, L.~Lavoura and M.~Tanimoto,
  JHEP {\bf 0407}, 078 (2004). 

\bibitem{Frigerio:2004jg}
  M.~Frigerio, S.~Kaneko, E.~Ma and M.~Tanimoto,
  Phys.\ Rev.\ D {\bf 71}, 011901 (2005). 


\bibitem{seidl}G.~Seidl,
  hep-ph/0301044.




\bibitem{Ma:2004br}See for instance 
P.~H.~Frampton and T.~W.~Kephart,
  Int.\ J.\ Mod.\ Phys.\ A {\bf 10}, 4689 (1995); 
C.~D.~Carone and R.~F.~Lebed,
  Phys.\ Rev.\ D {\bf 60}, 096002 (1999); 
P.~H.~Frampton and A.~Rasin,
Phys.\ Lett.\ B {\bf 478}, 424 (2000); 
  E.~Ma,
  hep-ph/0409288; 
K.~S.~Babu and J.~Kubo,
hep-ph/0411226; 
 J.~Kubo {\it et al.},  
  Prog.\ Theor.\ Phys.\  {\bf 109}, 795 (2003); 
  E.~Ma,
  Phys.\ Rev.\ D {\bf 61}, 033012 (2000); 
  S.~L.~Chen, M.~Frigerio and E.~Ma,
  Phys.\ Rev.\ D {\bf 70}, 073008 (2004)
  [Erratum-ibid.\ D {\bf 70}, 079905 (2004)]; 
T.~Araki, J.~Kubo and E.~A.~Paschos,
hep-ph/0502164.

\bibitem{books}J.~S.~Lomont,
  ``Applications of Finite Groups'',
  Acad. Press (1959) 346 p; 
  M.~Hamermesh,
  ``Group Theory and Its Application to Physical Problems'',
  Reading, Mass.: Addison-Wesley (1962) 509 p. 




\end{thebibliography}
\end{document}